\documentclass[12pt,letter]{article}
\usepackage{graphics}
\usepackage{graphicx}
\usepackage[latin1]{inputenc}
\usepackage{amsmath}
\usepackage{amssymb}

\newcommand{\Lyx}{L\kern-.1667em\lower.25em\hbox{y}\kern-.125emX\spacefactor1000}

\begin{document}
\bibliographystyle{plain} 
\pagestyle{plain} 
\pagenumbering{arabic}
\title{Crossover Between Organized and Disorganized States In Some Non-Equilibrium Systems}
\author {Diego Luis Gonz\'alez\footnote{die-gon1@uniandes.edu.co}\\
        Gabriel T\'ellez\footnote{gtellez@uniandes.edu.co}\\
        Departamento de F\'{\i}sica, Universidad de Los Andes\\
        A.~A.~4976 Bogot\'a, Colombia.}

\date{}

\maketitle

\begin{abstract}
We study numerically the crossover between organized and disorganized states of three non-equilibrium systems: the Poisson/coalesce random walk (PCRW), a one-dimensional spin system and a quasi one-dimensional lattice gas. In all cases, we describe this crossover in terms of the average spacing between particles/domain borders $\left\langle S(t)\right\rangle$ and the spacing distribution functions $p^{(n)}(s)$. The nature of the crossover is not the same for all systems, however, we found that for all systems the nearest neighbor distribution $p^{(0)}(s)$ is well fitted by the Berry-Robnik model. The destruction of the level repulsion in the crossover between organized an disorganized states is present in all systems. Additionally, we found that the correlations between domains in the gas and spin systems are not strong and can be neglected in a first approximation but for the PCRW the correlations between particles must be taken into account. To find $p^{(n)}(s)$ with $n>1$, we propose two different analytical models based on the Berry-Robnik model. Our models give us a good approximation for the statistical behavior of these systems in their crossover and allow us to quantify the degree of order/disorder of the system.
\end{abstract} 



\section{Introduction}
Many one-dimensional non-equilibrium systems exhibit a crossover between disorganized and organized states. This crossover usually depends on the value of one or more parameters which set the system in one of these states. Our main objective is to study the statistical behavior of three non-equilibrium systems in their crossovers between organized and disorganized states. The first system is the Poisson/coalesce random walk (PCRW) where the particles describe independent random walks and when two particles meet they coalesce with probability $k$, otherwise, they interchange their positions. The second system is a quasi one-dimensional gas, where the particles interact only by volume exclusion in presence of an external field. The last system is a one-dimensional spin lattice, where the particles interact by a coupling force $J$ in presence of an external driving field.

We propose two analytical models for the spacing distribution functions of these systems and compare them with the numerical results from the simulation. Our analytical models are based on the Berry-Robnik model introduced in Ref.~\cite{berry}. This model is used in quantum systems which are neither integrable nor fully chaotic in order to find an analytical approximation to the nearest neighbor distribution of the energy levels of the system, see Refs.~\cite{berry,amiet,lopac}. The Berry-Robnik model depends on one parameter which controls the crossover. The crossover is described through the nearest neighbor distribution $p^{(0)}(s)$, which gives us the probability that the distance between two consecutive levels is $s$. We complement the model proposed in Ref.~\cite{berry} calculating analytically the higher order spacing distribution functions $p^{(n)}(s)$ for $n>0$ and the pair correlation function $g(r)$ by two methods.

This paper is organized as follows. Because of its importance, in the second section we explain in detail the Berry-Robnik model. In third, fourth and fifth sections, we test our analytical models with three non-systems: the Poisson/coalesce random walk, the quasi one-dimensional gas and one-dimensional spin lattice respectively. These systems are also explained in those sections.

\section{Berry-Robnik model as a non-equilibrium system}
We consider a continuous one-dimensional ring with two kinds of particles $A$ and $B$ and normalized densities $\rho_A$ and $\rho_B$ respectively. The $A$ particles are subject to the reaction $A+A\rightarrow A$, i.e., they are coalescing random walkers. Particles $B$ describe independent random walks and they are subject to the reaction $B\rightarrow 0$. The amount of $B$ particles that disappear is calculated in such way that the number of $A$ particles, $N_A$, and the number of $B$ particles, $N_B$, satisfy the relation $N_B/N_A=\mathrm{constant}$ at each time step. This is the only interaction between both species of particles. 

For this system, $P^{(n)}(S,t)$ is the probability that the distance between two particles is $S$ at time $t$ under the condition that between these particles there are $n$ additional particles. When the average distance between neighbor particles is much smaller than the size of the lattice, the system exhibits a dynamical scaling for all values of $\rho_A$ and $\rho_B$. In this regime the spacing distributions can be scaled by using the variable change $s=S(t)/\left\langle S(t)\right\rangle$. Then, we obtain the time independent spacing distributions $p^{(n)}(s)$. 

It is known that in the scaling regime $N_A= L/(2\pi D t)^{1/2}$ where $D$ is the diffusion constant and $L$ is the size of the lattice, see Ref.~\cite{ben} for more information about the coalescing random walk (CRW). It is straightforward to show that $N_B=N_A \rho_B/(1-\rho_B)$, then, the average length between nearest neighbors satisfy $\left\langle S(t)\right\rangle=(1-\rho_B)(2 \pi D t )^{1/2}$.

The interaction between species warrants that the quotient between normalized densities $\rho_A$ and $\rho_B$ is constant for all time, i.e., the rate of disappearance of particles $B$ is proportional to the one for $A$ particles and the proportionality constant is $\rho_B/\rho_A$. Taking this into account, in the scaling regime, the system is the uncorrelated superposition of two independent systems of particles $A$ and $B$ with constant normalized densities. Our objective is to calculate the nearest neighbor distribution $p^{(0)}(s)$ of the whole system. This can be done easily because we have the superposition of two uncorrelated distribution functions $p^{(0)}_{i}(s)$ which satisfies 
\begin{equation}
\left\langle s_{i}\right\rangle=\int^{\infty}_{0} ds\,s\,p^{(0)}_{i}(s)=\frac{1}{\rho_{i}},
\end{equation} 
and
\begin{equation}
\int^{\infty}_{0} ds\,p^{(0)}_{i}(s)=1,
\end{equation} 
where $i=A\,\mathrm{or}\,B$ and $\rho_{i}$ is the density of particles corresponding to the systems $A$ or $B$. As is natural, the normalized densities satisfy $\rho_A+\rho_B=1$.

Let $E^{(1)}(s)$ be the probability to choose randomly an empty segment of length $s$. This probability is related to $p^{(0)}(s)$. In order to prove that, we consider the probability that there are no particles in the interval $q$ to $q+r$, given that there is a particle at $q$. This probability is given by $\int^{\infty}_{r}p^{(0)}(x)\,dx$.

In our case we have the uncorrelated superposition of two kinds of particles, then we can write
\begin{equation}\label{eqp}
\int^{\infty}_{r}p^{(0)}(x)dx=\rho_A\,q_B(r)\int^{\infty}_{r}p^{(0)}_{A}(y)dy +\rho_B\,q_A(r) \int^{\infty}_{r}p^{(0)}_{B}(y)dy ,
\end{equation}
where, for example, $\rho_A$ is the probability that the particle in position $q$ belongs to the $A$ system, $\int^{\infty}_{r}p^{(0)}_{A}(x)dx$ and $q_B(r)$ are the probabilities that there are no particles of the systems $A$ and $B$ in the interval, respectively. In order to find $q_B(r)$, we choose a particle randomly, the probability that the chosen point $q$ lies within a gap of length $\sigma$ to $\sigma+d\sigma$ is proportional to $\sigma\,p^{\left(0\right)}_B(\sigma)$, to normalize this equation we use the fact that $\left\langle s_B\right\rangle=1/\rho_B$, obtaining the probability distribution $\rho_B\,d \sigma\,\sigma\,p^{\left(0\right)}_B(\sigma)$. Now, the probability that the distance to the next particle is $r$, given that the point is in the gap of length $\sigma$, is zero if $\sigma<r$ and $1/\sigma$ if $1\,\leq\,r\leq\,\sigma$. The probability of not having a particle of the system $B$ in an interval of length $r$ is $(1-r/\sigma)\theta(\sigma-r)$ ($\theta$ is the unit step function).

Thus, the unconditional probability that the distance until the next particle is $r$ is given by 

\begin{equation}
q_B(r)=\rho_B\int^{\infty}_{r}p^{(0)}_B(\sigma)(\sigma-r)d\sigma.
\end{equation}
The same argument can be done for the second term of equation (\ref{eqp}). Thus Eq.~(\ref{eqp}) takes the form
\begin{eqnarray}\label{eqz}
\int^{\infty}_{r}p^{(0)}(x)dx &=&\rho_A \int^{\infty}_{r}p^{(0)}_{A}(y) dy\rho_B\int^{\infty}_{r}p^{(0)}_B(\sigma)(\sigma-r)d\sigma \nonumber\\
&+&\rho_B \int^{\infty}_{r}p^{(0)}_{B}(y) dy \rho_A\int^{\infty}_{r}p^{(0)}_A(\sigma)(\sigma-r)d\sigma.
\end{eqnarray}
The probability $E^{(1)}(s)$ is the probability that $r>s$, then, integrating Eq.~(\ref{eqz}) over $r$ we find
\begin{equation}\label{e0}
E^{(1)}(s)=\int^{\infty}_{s}dr\int^{\infty}_{r}dx p^{(0)}(x)=\rho_A E^{(1)}_A(s)\rho_B E^{(1)}_B(s),
\end{equation}
where we define
\begin{equation}\label{e00}
E^{(1)}_i(s)=\int^{\infty}_{s}d\sigma p_i^{(0)}(\sigma)(\sigma-s)=\int^{\infty}_{s}d\sigma\int^{\infty}_{\sigma}dx p_i^{(0)}(x).
\end{equation} 

From equations (\ref{e0}) and (\ref{e00}), the spacing distribution function $p^{(0)}(s)$, which results of the mix of systems $A$ and $B$ is given by
\begin{equation}
p^{(0)}(s)=\frac{d^2E^{(1)}(s)}{ds^2}.
\end{equation} 

The fact that $E^{(1)}(s)$ is given by the product of $E^{(1)}_i(s)$ functions is natural, because we used two statistical independent sequences. It is easy to prove that the $E^{(1)}_i(s)$ functions satisfy 
\begin{equation}
E^{(1)}_i(0)=\frac{1}{\rho_i},
\end{equation}
and
\begin{equation}
\frac{dE^{(1)}_i(0)}{ds}=-1.
\end{equation}
From the above equations it follows that $p^{(0)}(0)$ is given by
\begin{equation}
p^{(0)}(0)=1+\rho_A\left(p^{(0)}_A(0)-\rho_A\right)+\rho_B\left(p^{(0)}_B(0)-\rho_B\right).
\end{equation}
Then, even if $p_A^{(0)}(0)=0$ and $p_B^{(0)}(0)=0$ it can happen that $p^{(0)}(0)\neq0$.
In our particular case, we superpose one Wigner distribution 
\begin{equation}
p^{(0)}(s,\rho_A)=\frac{\pi}{2}\,\rho_A^2 s\,\mathrm{exp}[-\frac{\pi}{4}\,\rho_A^2 s^2]
\end{equation}
with density $\rho_A$ and one Poisson distribution 
\begin{equation}
p^{(0)}(s,\rho_B)=\rho_B \mathrm{exp}[-\rho_B\, s]
\end{equation}
with density $\rho_B$ , then, we have
\begin{eqnarray}\label{berryeq}
p^{(0)}(s,q)=e^{-q s}\left(q^2\mathrm{erfc}\left(\frac{\sqrt{\pi}}{2}(1-q)s\right)+\left(2 q(1-q)+\frac{\pi}{2}(1-q)^3 s\right)e^{-\frac{\pi}{4}(1-q)^2 s^2}\right),
\end{eqnarray}
where $\rho_B\equiv q$ is the density of Poisson sequence, $\rho_A=1-q$ is the one for the Wigner sequence and $\mathrm{erfc}(z)$ is the complementary Gaussian error function. Note that Eq.~(\ref{berryeq}) reduces to the Poisson distribution for $q=1$, and for $q=0$ it reduces to the Wigner distribution.
This result is well-know in random matrices theory and corresponds to the Berry-Robnik model for crossover between chaotic and non-chaotic behavior in quantum systems, see Ref.~\cite{berry}. 

\subsection{Higher spacing distribution functions}
In Refs.~\cite{gonzalez,gonzalez3} the authors show that the nearest neighbor distribution is not enough to describe the complete statistical behavior of a non-equilibrium system, because several different systems could share the same nearest neighbor distribution. Because of that, we generalize the Berry-Robnik model for higher spacing distribution functions. Let $E^{\left(n\right)}(x_1,y_1,\cdots,x_n,y_n)$ be the joint probability that the intervals $[x_i,y_i]$ $(i=1,2,\cdots,n)$ are empty. The intervals are non overlapping and ordered $x_1<y_1<\cdots<x_n<y_n$. Because of the independent superposition nature of the Berry-Robnik model, $E^{\left(n\right)}(x_1,y_1,\cdots,x_n,y_n)$ can be written as
\begin{equation}\label{enxy}
E^{\left(n\right)}(x_1,y_1,\cdots,x_n,y_n)=\rho_A^{n}E_A^{(n)}(x_1,y_1,\cdots,x_n,y_n)\rho_B^{n}E_B^{(n)}(x_1,y_1,\cdots,x_n,y_n),
\end{equation} 
where $E_A^{(n)}(x_1,y_1,\cdots,x_n,y_n)$ and $E_B^{(n)}(x_1,y_1,\cdots,x_n,y_n)$ are the joint probabilities for particles $A$ and $B$ respectively. 
The authors of Ref.~\cite{ben2}, showed that spacing distribution functions $p^{(n)}(s)$ can be calculated from Eq.~(\ref{enxy}), let us summarize their most important result. Let $p^{\left(n\right)}(s)$ be the probability that given one particle its $(n+1)$-th neighbor is at a distance $s$. From its definition $p^{\left(n\right)}(s)$ is given by
\begin{equation}\label{pn}
p^{\left(n\right)}(s)=\int_{0<y_1<\cdots<y_n<s}\omega^{\left(n+2\right)}(0,y_1,\cdots,y_n,s)dy_1 \cdots dy_n,
\end{equation}
with
\begin{equation}\label{omega}
\omega^{\left(n\right)}(x_1,\cdots,x_n)=-\left.\frac{\partial^n E^{\left(n-1\right)}(x_1,y_1,\cdots,x_{n-1},y_{n-1})}{\partial x_1\cdots \partial x_{n-1}\partial y_{n-1}}\right|_{y_1=x_2,\cdots,\,y_{n-1}=x_n}.
\end{equation}

On the other hand, it is well-know that for CRW, $E_A^{(n)}(x_1,y_1,\cdots,x_n,y_n)$ is given by~\cite{ben}
\begin{equation}\label{ena}
E_A^{(n)}(x_1,y_1,\cdots,x_n,y_n)=\sum_{p}\sigma_p E_A^{(1)}(z_{1,p},z_{2,p})\cdots E_A^{(1)}(z_{2n-1,p},z_{2n,p}),
\end{equation}
where $z_{1,p}, z_{2,p}, . . . , z_{2n,p}$ symbolize an ordered permutation, $p$, of the variables $x_1, y_1, . . . , x_n, y_n$, such that

\begin{equation}
z_{1,p} < z_{2,p}, z_{3,p} < z_{4,p},\cdots, z_{2n-1,p} < z_{2n,p},
\end{equation}
\begin{equation}
z_{1,p} < z_{3,p} < z_{5,p}< \cdots < z_{2n-1,p}.
\end{equation}
In Eq.~(\ref{ena}) $\sigma_p$ is the signature of the permutation, i.e., $\sigma_p$=1 for even permutations and $\sigma_p=-1$ for odd permutations.

The function $E_A^{(1)}(x_1,y_1)$ is the probability that from $x_1$ to $y_1$ the lattice is empty. Then it is possible generate the complete solution for the CRW from $E_A^{(1)}(x_1,y_1)$, which is given by the solution of the diffusion equation under the suitable boundary conditions (see Ref.~\cite{ben}). In fact, the exact expression for this function is 
\begin{equation}\label{e1}
E_A^{(1)}(x_1,y_1)=\frac{1}{\rho_A}\mathrm{erfc}\left(\frac{\sqrt{\pi}\rho_A}{2}(y_1-x_1)\right),
\end{equation}
for additional information see Refs.~\cite{ben} and \cite{gonzalez2}. The particles $B$ describe independent random walks, then, we have
\begin{equation}
E_B^{(1)}(x_1,y_1)=\frac{1}{\rho_B}e^{-\rho_B (y_1-x_1)},
\end{equation}
and
\begin{equation}\label{enb}
E_B^{(n)}(x_1,y_1,\cdots,x_n,y_n)=\prod^{n}_{i=1}\frac{1}{\rho^{n}_B}e^{-\rho_B (y_i-x_i)}.
\end{equation} 
From equations (\ref{enxy}), (\ref{ena}), (\ref{e1}) and (\ref{enb}) we can calculate $E^{(n)}(x_1,y_1,\cdots,x_n,y_n)$ for the whole system and then, we can use (\ref{pn}) and (\ref{omega}) to calculate $p^{(n)}(s)$. By using this formalism it is also possible to calculate the $n$-point correlation function
\begin{equation}\label{npoints}
\rho^{\left(n\right)}(x_1,\cdots,x_n)=(-1)^n \frac{\partial^n}{\partial y_1\cdots\partial y_n}\left.E^{\left(n\right)}(x_1,y_1,\cdots,x_n,y_n)\right|_{y_1=x_1,\cdots,\,y_n=x_n},
\end{equation}
in fact, the pair correlation function is given by
\begin{equation}
g(r)=1-(1-q)^2e^{-\frac{\pi}{2}(1-q)^2r^2} +\frac{\pi}{2}e^{-\frac{\pi}{4}(1-q)^2 r^2}(1-q)^3 r\,\mathrm{erfc}\left(\frac{\sqrt{\pi}}{2}(1-q)r\right).
\end{equation}
The above equation takes the form $g(r)=1$ for $q=1$. For $q=0$ we recover the pair correlation function of the CRW, see Ref~\cite{ben}.

\section{A simple model of crossover between organized and disorganized states: The Poisson/coalesce random walk}
Consider a one-dimensional ring with $L$ sites and $n_p$ particles, then, the particle density is given by $\rho=n_p/L$. The particles describe independent random walks and when two particles meet they coalesce with probability $k$, otherwise, they interchange their positions. This system was studied previously in Refs.~\cite{ben0,priv}. The algorithm used in the simulation is the following:
\begin{enumerate}
	\item $n_p$ particles are randomly inserted in a $L$ sites lattice.
	\item A particle is chosen at random.
	\item The particle can move to the left or to the right with same probability. If the particle is moved to an occupied site, the particles coalesce with probability $k$ or they interchange their positions with probability $1-k$.
	\item In a time unit all particles are moved.
\end{enumerate}

In the limit $t\gg1$, finite systems reach a non-equilibrium steady state (NESS), where there is only one particle which executes a simple random walk and $\left\langle S(t)\right\rangle=L$. As we will see soon, for an infinite size system in the same limit, the average length between nearest neighbor particles grows as $t^{1/2}$ for $k>0$ and the system is statistically equivalent to the CRW.  

\begin{figure}[htp]
\begin{center}
\includegraphics[scale=0.8]{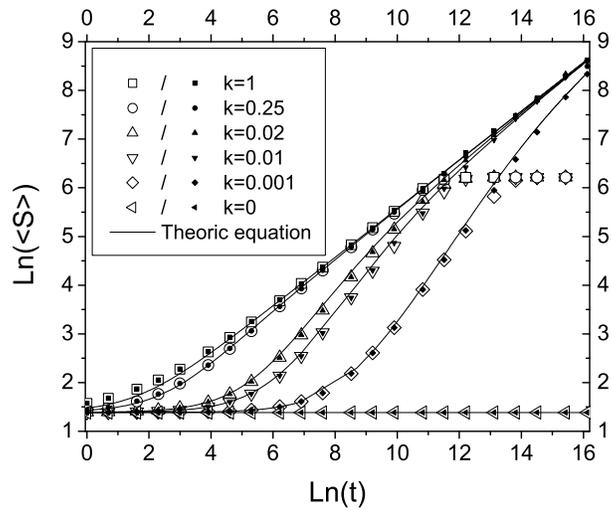}
\end{center}
\caption{Behavior of $\left\langle S\right\rangle$ as function of $t$. In the simulation we took two lattices with $L=10000$ (filled symbols) and $L=500$ (empty symbols) over $500$ realizations. In both cases the initial density was $1/4$. The continuous line represents Eq.~(\ref{cdet}).}
\label{pcrwxprom}
\end{figure}

It is possible to derive an approximate analytical solution by using the inter-particle distribution function method (IPDF), see Refs.~\cite{ben,ben0}. Let $E_n$ be the probability to find an empty segment of length $n$ in the lattice. The master equation for $E_n$ can be written as
\begin{equation}\label{en}
\frac{\partial E_n(t)}{\partial t}=\frac{2 D}{\Delta x^2}\left(E_{n+1}(t)-2E_n(t)+E_{n-1}(t)\right)+\frac{2D}{\Delta x^2}\left(k-1\right)P(\overbrace{\circ\cdots\circ}^{n-1}\bullet\bullet),
\end{equation}
with boundary conditions $E_0(t)=1$ and $E_\infty(t)=0$. The probability $P(\overbrace{\circ\cdots\circ}^{n-1}\bullet\bullet)$ cannot be written in terms of $E_n(t)$. However in Refs.~\cite{ben,ben0} the authors propose an approximation for this probability
\begin{equation}
P(\overbrace{\circ\cdots\circ}^{n-1}\bullet\bullet)\approx\frac{P(\overbrace{\circ\cdots\circ}^{n-1}\bullet)P(\bullet\bullet)}{P(\bullet)}
\end{equation}
which can be written in terms of $E_n(t)$ as
\begin{equation}\label{papp}
P(\overbrace{\circ\cdots\circ}^{n-1}\bullet\bullet)\approx\frac{(1-2E_1+E_2)(E_{n-1}-E_n)}{1-E_1}.
\end{equation}
This approximation allows us to calculate the concentration of particles $c(t)=1/\left\langle S(t)\right\rangle$. From Eqs.~(\ref{en}) and (\ref{papp}) and summing over the index $n$ it is easy to find 
\begin{equation}
\sum^{\infty}_{n=1}\frac{\partial E_n(t)}{\partial t}=\frac{2 D}{\Delta x^2}(E_0(t)-E_1(t))-\frac{2 D}{\Delta x^2}(1-k)\frac{(1-2E_1(t)+E_2(t))}{1-E_1(t)}E_{0}(t).
\end{equation}
Taking into account that $\partial_t E_1(t)=2 D k (1-2E_1+E_2)/\Delta x^2$ and making the approximation $\sum^{\infty}_{n=1}E_n(t)\approx (2/\pi)/ (1-E_1(t))$, in Ref~\cite{ben0} it is found that
\begin{equation}\label{dcdt}
\frac{2}{\pi}\frac{d}{d\tau}\left(\frac{1}{c(\tau)}\right)=c(\tau)+\frac{(1-k)}{k\,c(\tau)}\frac{d c(\tau)}{d\tau}
\end{equation}
where $\tau=2 D t/\Delta x^2$ and $c(\tau)=1-E_1(\tau)$. Equation (\ref{dcdt}) can be integrated obtaining
\begin{equation}\label{cdet}
c(\tau)=\frac{2 c_0^2 k}{c_0^2(k-1)\pi+\sqrt{c_0^2(-2k+c_0(k-1)\pi)^2+4 c_0^4k^2\pi\tau}},
\end{equation}
with $c_0\equiv c(0)$. Note that for $\tau\rightarrow \infty$, we have $c(\tau)\propto\tau^{-1/2}$ as we mentioned above. In figure \ref{pcrwxprom}, we show the behavior of $\left\langle S(t)\right\rangle=1/c(\tau)$ for different values of $k$, this result was first shown in Ref.~\cite{ben0}. The agreement between Eq.~(\ref{cdet}) and the simulation is very good. Because of the finite size effects in the simulation, for $L=500$, the system does not reach the asymptotic exponent $\beta=1/2$ $(\left\langle S(t)\right\rangle\propto t^{\beta})$ but for $L=10000$ we can see this regime for $k=1$, $k=0.25$, $k=0.02$ and $k=0.01$. For low values of $k$, the time that the system remains disorganized with $\left\langle S(t)\right\rangle$ almost constant, is larger than for large values of $k$. 

The nearest neighbor distribution $p^{(0)}(s)$ evolves in the following way. The system starts in an disorganized state in such way that $p^{(0)}(s)$ is described by the Poisson distribution, then, for $k\neq0$ the system evolves and $p^{(0)}(s)$ is deformed continuously until it reaches the Wigner distribution, for large systems. For small systems, the finite size effects appear before this regime is attained, i.e., small systems reaches the NESS without reaching the scaling regime. In the $k=0$ case, the system remains disorganized for all values of $t$.    

In the continuum limit, Eq.~(\ref{en}) takes the form   
\begin{equation}\label{ext}
\frac{\partial E^{(1)}(x,t)}{\partial t}=2 D\frac{\partial^{2} E^{(1)}(x,t)}{\partial x^2}+2 D(k-1) \frac{\partial E^{(1)}(x,t)}{\partial x}\frac{\left.\frac{\partial^2 E^{(1)}(x,t)}{\partial x^2}\right|_{x=0}}{\left.\frac{\partial E^{(1)}(x,t)}{\partial x}\right|_{x=0}}, 
\end{equation}
with boundary conditions $E^{(1)}(0,t)=1$ and $E^{(1)}(\infty,t)=0$. This equation can be solved easily for $k=1$. For arbitrary values of $k$, Eq.~(\ref{ext}) cannot be solved exactly, however some approximate expressions were found in Refs.~\cite{ben,ben0,priv}, but they involve self-consistent forms which are difficult to handle. Motivated by this fact, we propose to use the Berry-Robnik model to find an approximate analytical solution to the statistical behavior of this system. 

\begin{figure}[htp]
\begin{center}
$\begin{array}{cc}
\includegraphics[scale=0.55]{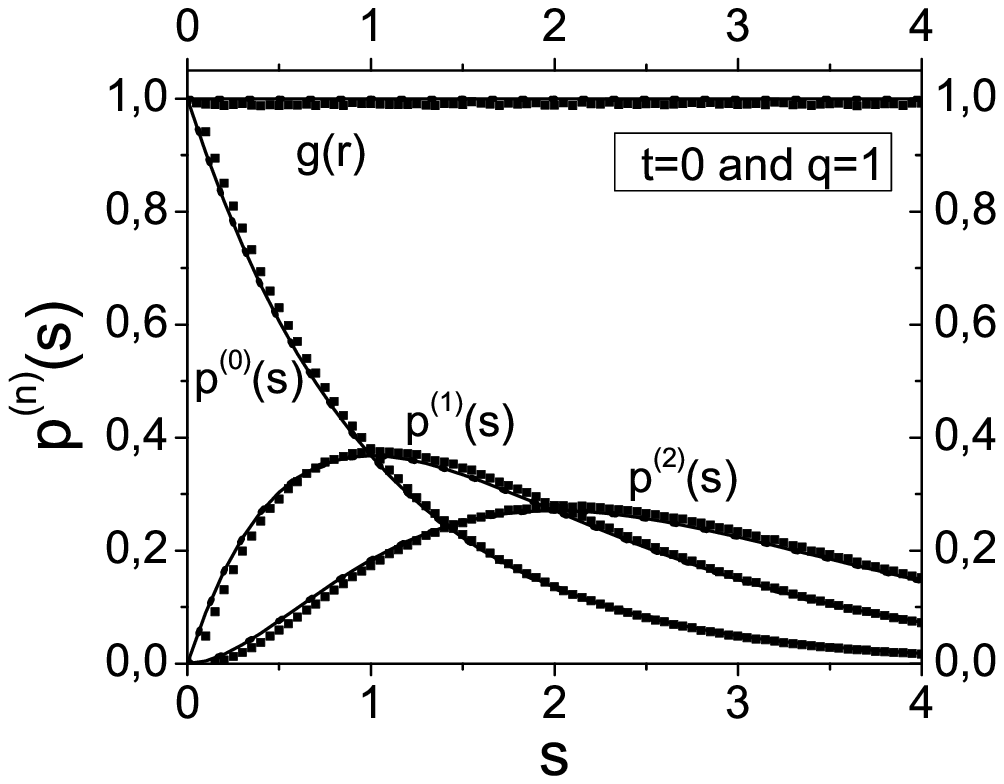}&
\includegraphics[scale=0.55]{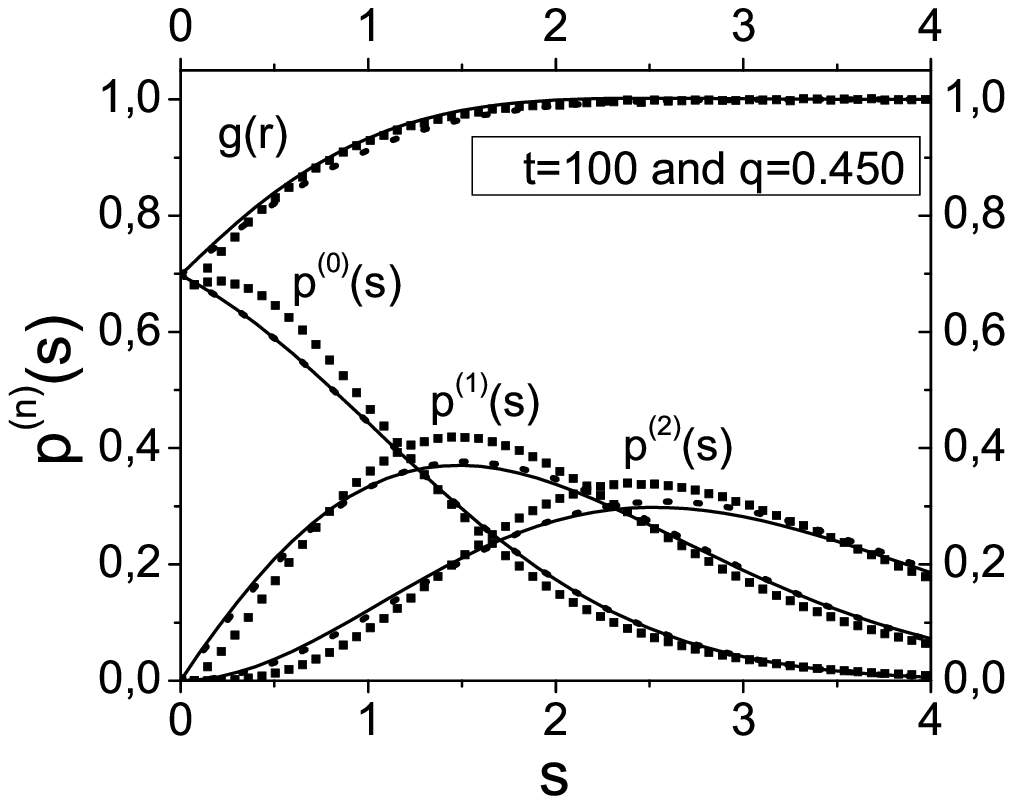}\\
(a) & (b) \\
\includegraphics[scale=0.55]{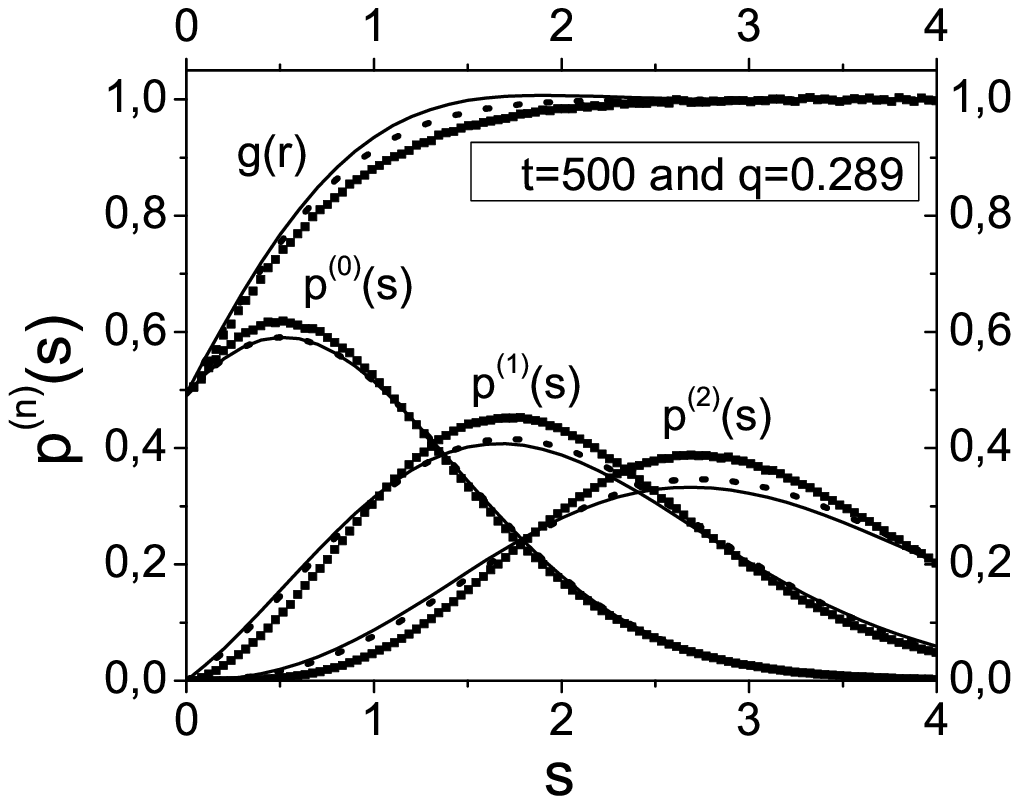}&
\includegraphics[scale=0.55]{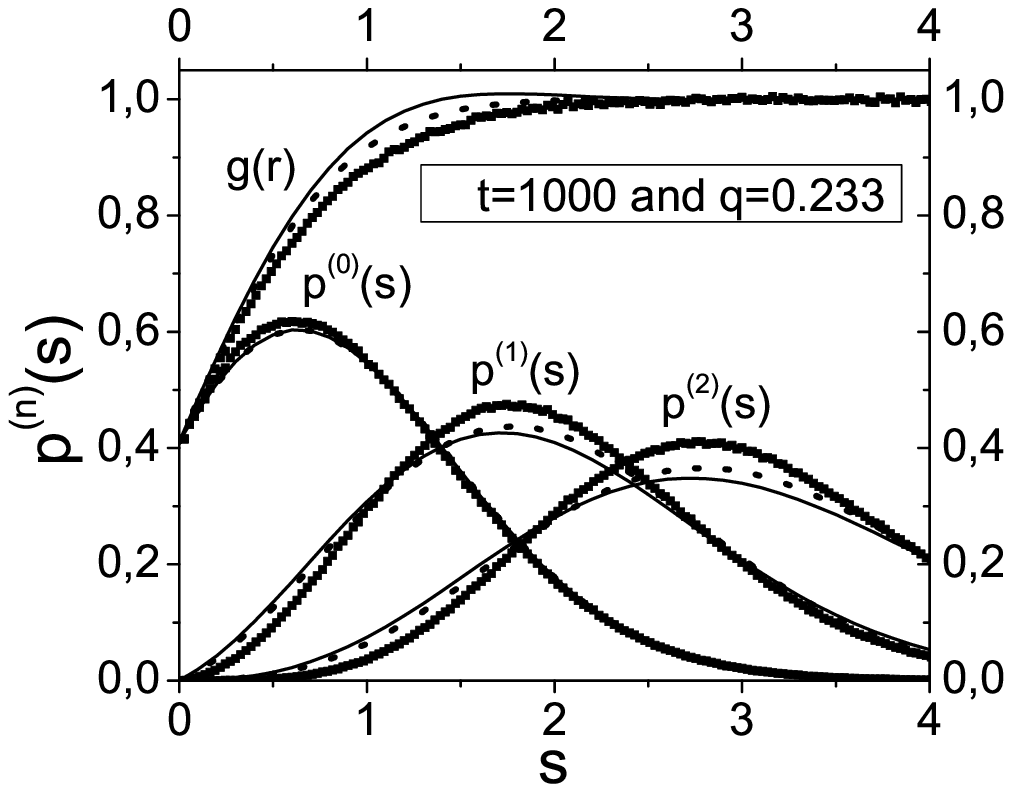}\\
(c) & (d) \\
\includegraphics[scale=0.55]{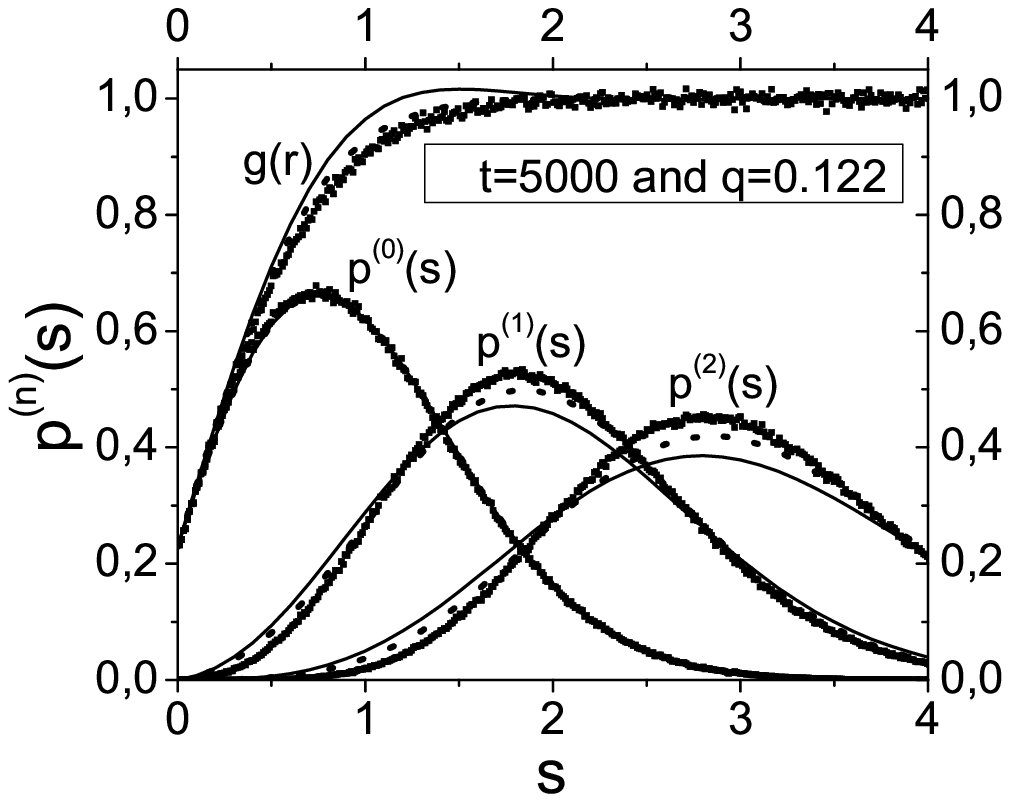}&
\includegraphics[scale=0.55]{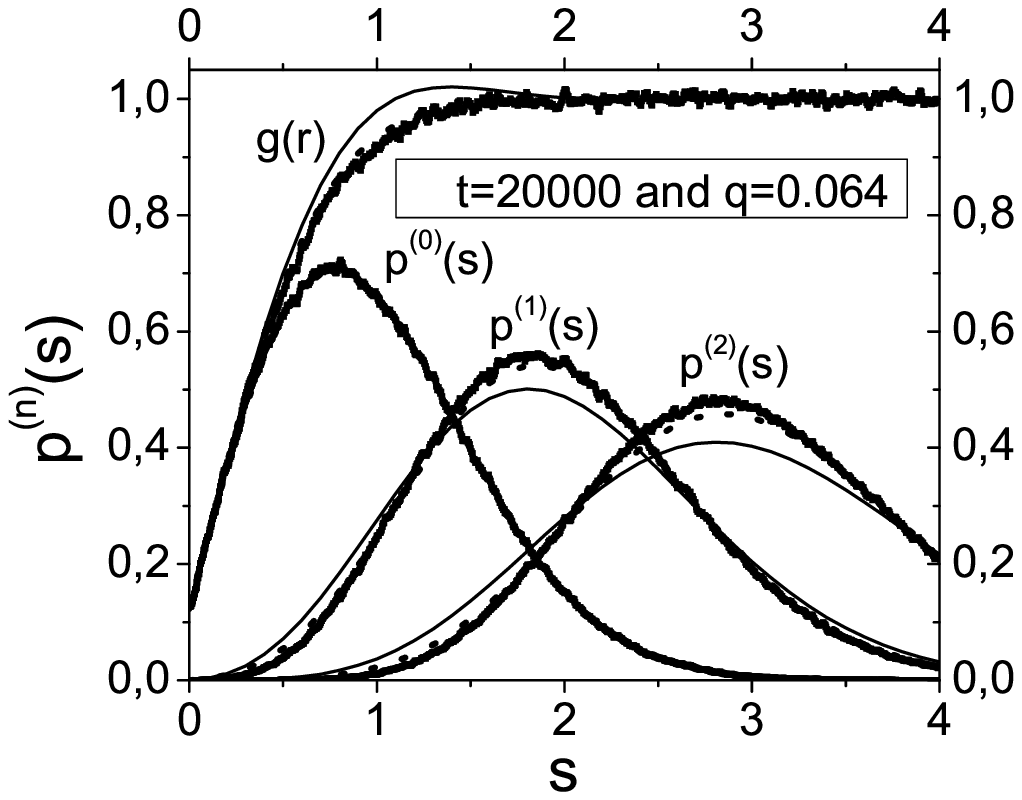}\\
(e) & (f) \\ 
\end{array}$
\end{center}
\caption{Crossover between the Poisson and Wigner distribution for the PCRW, for $k=0.05$, for various times $t$. The continuous line corresponds to the BR+IIA model, the dash line corresponds to the GBR model and the square dots correspond to the simulation.}
\label{pcwr}
\end{figure}

Equation (\ref{berryeq}) provides a good fit for the crossover in this reaction-diffusion model during its time evolution as we can see in figure \ref{pcwr}. For the nearest neighbor distribution the fit is almost perfect for low and high values of $s$, but for intermediate values we can see little differences in the interval $0.25<s<1$ for intermediate values of $k$. In this figure all data was taken at different times with $k=0.05$ over $20000$ realizations, the initial density was $0.1$. The fit parameter is $q$. We found the appropriate value of $q$ by using the numerical results for $p^{(0)}(s)$, equation (\ref{berryeq}) and the minimum square criteria \footnote{This also can be done for values of $k$ near to $1$ or long times by using the numerical value of $p^{(0)}(0)$ obtained from the simulation and taking into account that $p^{(0)}(0)=2 q-q^2$ for the Berry-Robnik model.}. 

The higher spacing distributions have been calculated with two methods. In the first one, we use our extension for the original Berry-Robnik Eq.~(\ref{enxy}) with equation (\ref{pn}). From now on, we call it the generalized Berry-Robnik (GBR) model. For the second one, we use equation (\ref{berryeq}) with the independent interval approximation (IIA). We call this method the Berry-Robnik+IIA (BR+IIA) model. In the IIA approximation, the entire statistical behavior of the system is described by the nearest neighbor distribution $p^{(0)}(s)$ and the joint probability to find the particles in positions $x_1,\cdots,x_N$ can be written as the product of independent product of the $N$ nearest neighbor distributions
\begin{equation}\label{pnxn}
P_N(x_1,\cdots,x_N)=\frac{1}{Z_N}p^{(0)}(x_2-x_1)\cdots p^{(0)}(x_N-x_{N-1})p^{(0)}(x_{1}+L-x_{N}),
\end{equation}
where the partition function $Z_N$ is the normalization constant
\begin{equation}
\label{eq:part-IIA}
Z_N=\int_{x_1<x_2<\ldots<x_N<x_1+L} dx_1\ldots dx_N\,
\prod^{N}_{i=1}p^{(0)}(x_{i+1}-x_i)\,.
\end{equation}
In the IIA approximation the correlations among intervals $[x_1,x_2]$, $\cdots$, $[x_{N-1},x_N]$ are neglected, for a review of this method see Ref.~\cite{gonzalez}. As is natural, the generalized Berry-Robnik model and the Berry-Robnik+IIA model are equivalents in the limit $q\rightarrow 1$ because in this regime those correlations are not strong and can be neglected. 

In general, for the PCRW the fits are better for the GBR model than
the BR+IIA. This means that for this system the information contained
in $p^{(0)}(s)$ is not enough to describe the entire statistical
behavior of the system, i.e., the correlation among intervals can not
be neglected as it happens in the BR+IIA model. These correlations can
be neglected only for small times $(\tau \leq \tau_1$), when the
particles do not coalesce enough to build strong correlations between
them. The value of $\tau_1$ as a function of $k$ can be computed from
equation (\ref{dcdt}) as follows. Integrating Eq. (\ref{dcdt}), it is
easy to find
\begin{equation}\label{tau1}
\tau=\frac{1-k}{k}\left(\left\langle S(\tau)\right\rangle-\left\langle S(0)\right\rangle\right)+\frac{1}{\pi}\left(\left\langle S(\tau)\right\rangle^2-\left\langle S(0)\right\rangle^2\right).
\end{equation}
Because of the reaction between particles, $\left\langle S(\tau)\right\rangle$ grows in time for $k>0$ making possible to write $\left\langle S(\tau)\right\rangle$ as $\left\langle S(0)\right\rangle$ plus an increment $\left\langle \Delta S\right\rangle$. In this way Eq. (\ref{tau1}) takes the form
\begin{equation}\label{tau1delta}
\tau=\frac{1-k}{k}\left\langle \Delta S\right\rangle+\frac{1}{\pi}\left\langle \Delta S\right\rangle^2+\frac{2}{\pi}\left\langle \Delta S\right\rangle \left\langle S(0)\right\rangle.
\end{equation}
The correlations between particles can be neglected for small times when $\left\langle \Delta S\right\rangle \ll 1$, then, we can neglect $\left\langle \Delta S\right\rangle^2$ in Eq. (\ref{tau1delta}). Finally if we consider that $\left\langle \Delta S\right\rangle$ must be a fraction $\epsilon$ of $\left\langle S(0)\right\rangle$, we find
\begin{equation}\label{tau1f}
\tau_1=\epsilon \left(\frac{1-k}{k}\left\langle S(0)\right\rangle+\frac{2}{\pi}\left\langle S(0)\right\rangle^2\right).
\end{equation}
This equation was derived first in Ref. \cite{ben0}. From Eq. (\ref{tau1f}), it is clear that there is a time where the interaction between particles goes unnoticed even in the case of $k=1$, where, $\tau_1$ is the typical time that one particle needs to reach one of its nearest neighbors. In the limit $k \rightarrow0$, we have $\tau_1 \rightarrow \infty$ as is natural because in this case $\left\langle S(\tau)\right\rangle$ is a constant.
  
We conclude, that, for the PCRW the spacing distribution and the pair correlation functions can be approximated from the uncorrelated superposition of a Poisson and a Wigner distribution functions as is proposed in the GBR model. 

\begin{figure}[htp]
\begin{center}
$\begin{array}{cc}
\includegraphics[scale=0.58]{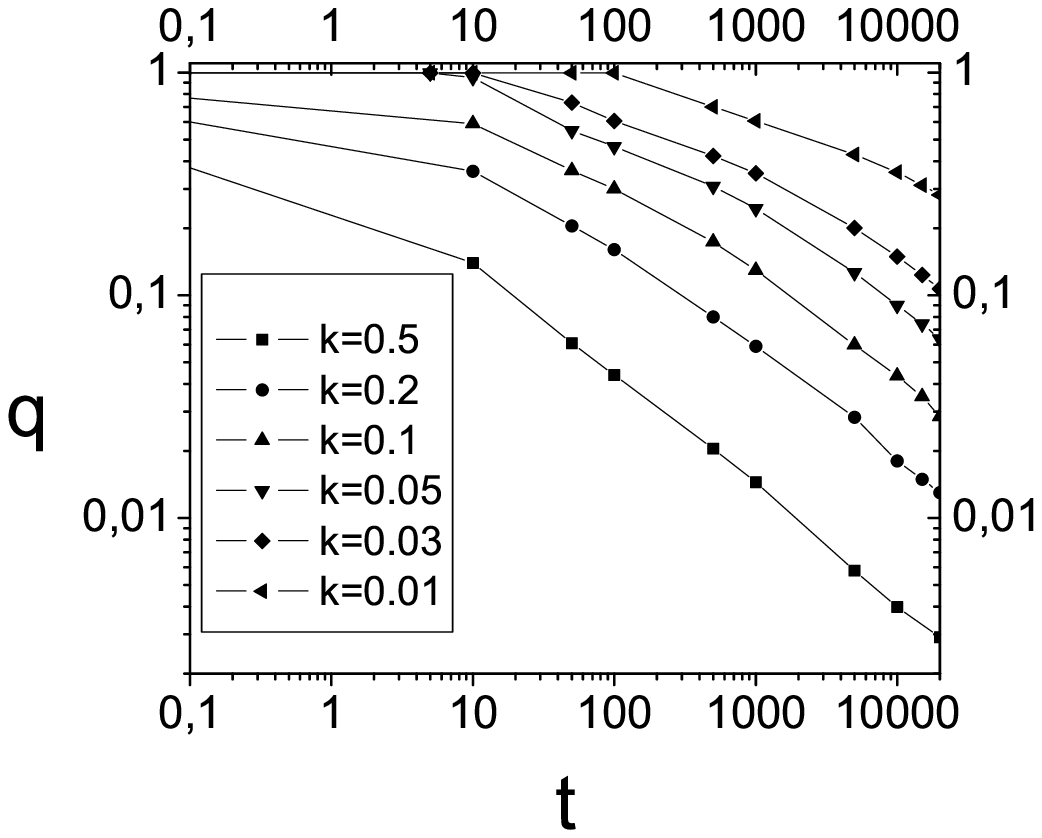}&
\includegraphics[scale=0.60]{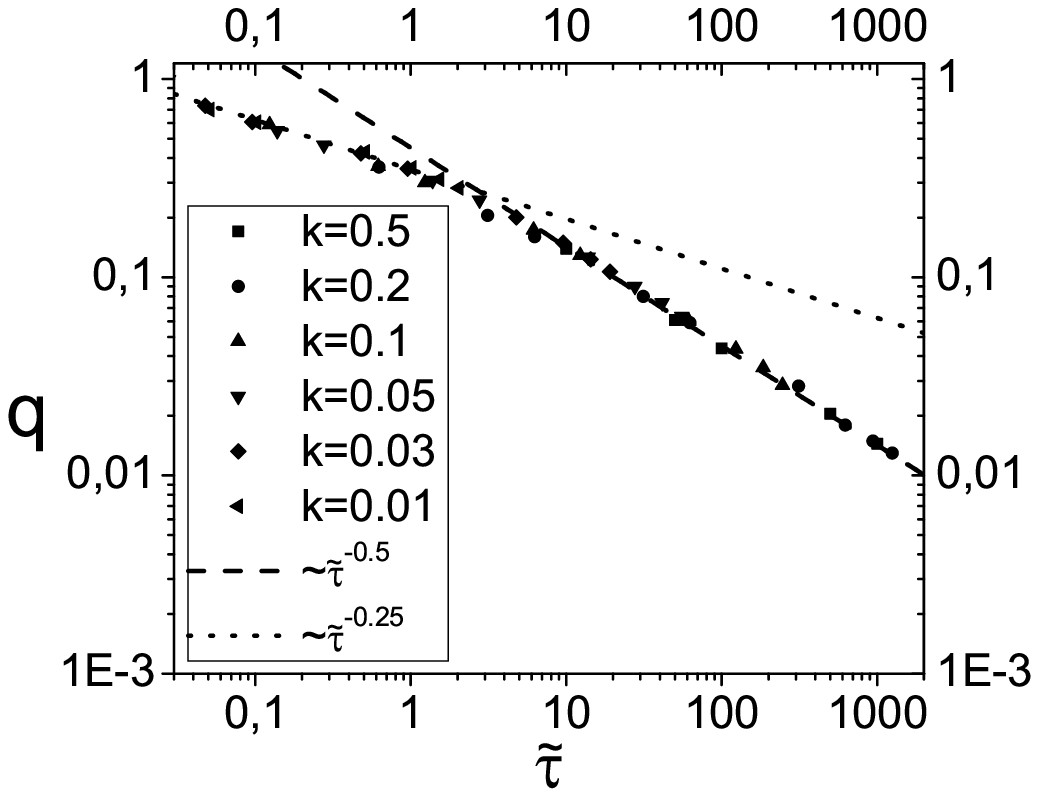}\\
(a) & (b) \\
\end{array}$
\end{center}
\caption{Behavior of the fit parameter $q$ as function of $t$ for different values of $k$.}
\label{pcrwt}
\end{figure}

\begin{figure}[htp]
\begin{center}
\includegraphics[scale=0.75]{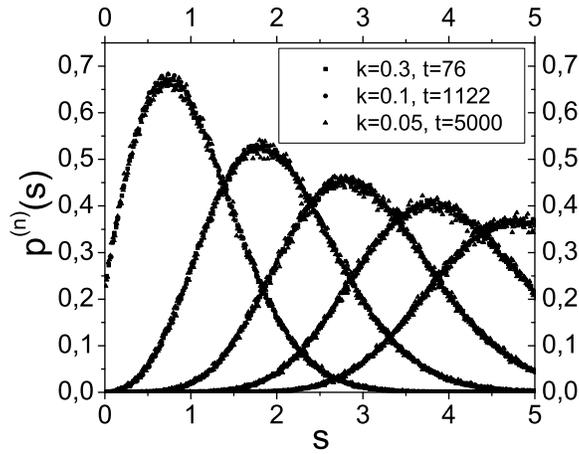}
\end{center}
\caption{Statistical behavior for the PCRW for different values of $k$ and $t$ with $\tau k^2/(1-k)^2\approx13.9$, in all cases $q\approx0.112$.}
\label{aver}
\end{figure}

An alternate way to analyze the crossover of this system, is to study the spacing distribution functions at a fixed given time, but for different values of $k$. In order to establish a connection between this picture for the transition and the one shown in figure \ref{pcwr}, it is necessary to find the correct combination of $k$ and $\tau$ which gives the same statistical behavior for different values of these parameters. To find it, in figure \ref{pcrwt}-(a) we show the behavior of $q$ as a function of $\tau$ for different values of $k$. We found that making the change of variable \begin{equation}\label{cv}
\tilde{\tau}=\tau k^2/(1-k)^2,
\end{equation} 
all lines shown in figure \ref{pcrwt}-(a) collapse in a single one,
see figure \ref{pcrwt}-(b). Then, different combinations of $k$ and
$\tau$ with fixed $\tau k^2/(1-k)^2$ give the same statistical
behavior, i.e., give the same value for the fit parameter $q$. This is
shown in figure \ref{aver}, the data was taken in all cases over
$20000$ realizations with $\tau k^2/(1-k)^2\approx13.9$. For large
values of $\tilde{\tau}$, $\tilde{\tau}\gg1$, we found that $q\propto
\tilde{\tau}^{-0.5}$ and for low values of $\tilde{\tau}$,
$\tilde{\tau}\ll1$, $q\propto
\tilde{\tau}^{-0.25}$.

The physical meaning of the change of variable (\ref{cv}) is clear if we introduce the crossover time, $\tau_2$, between the intermediate regime
(which starts when $\tau>\tau_1$) and the long-time regime (when $k$ renormalizes to 1). This time is estimated in Ref. \cite{ben0}, by expanding Eq.~(\ref{cdet}) in powers of $1/\sqrt{\tau}$,
\begin{equation}
\tau_2\propto\frac{(1-k)^2}{k^2},
\end{equation}   
which is precisely the scaling factor used in figure
\ref{pcrwt}-(b). The change of variable~(\ref{cv}) is
$\tilde{\tau}=\tau/\tau_2$.

\section{Crossover in the quasi one dimensional gas system}
This system was originally studied in \cite{mettetal}. There, the authors studied the
biased diffusion of two species in a fully periodic $2\times L$ rectangular lattice half filled with two types of particles labeled by their charge, there are $L/2$ particles with charge $+$ and $L/2$ particles with charge $-$. An infinite external field drives the two species in opposite directions along the $x$ axis (long axis). The only interaction between particles is an excluded volume constraint, i.e., each lattice site can be occupied by only one particle. The system evolves in time according to the following dynamical rules:

\begin{enumerate}
\item $L$ particles are randomly inserted in a $2\times L$ rectangular
  lattice, $\frac{L}{2}$ particles $(+)$ and $\frac{L}{2}$ particles
  $(-)$, the remaining sites are empty. Periodic boundary conditions
  are imposed in both directions of the lattice. Let the $x$ axis be
  the long axis of length $L$.
\item Two neighbor sites are chosen at random. The contents of the
  sites are exchanged with probability $1$ if the neighbor sites are
  particle-hole, but if they are particle-particle the contents are
  exchanged with probability $\gamma$. The exchanges which result in
  $+/-$ particles moving in the positive/negative $x$ direction are
  forbidden due to the action of the external field.
\item A time unit, corresponds to $2L$ attempts of exchange.
\end{enumerate}

\begin{figure}[htp]
\begin{center}
$\begin{array}{c}
\includegraphics[scale=0.4]{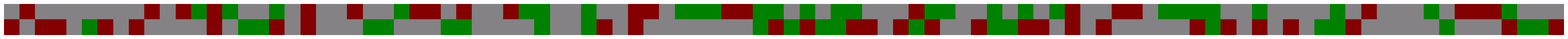}\\
(a)\\
\includegraphics[scale=0.4]{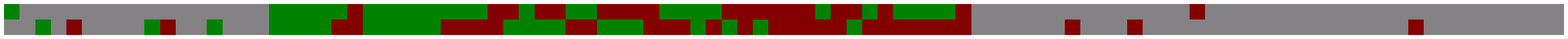}\\
(b)\\
\includegraphics[scale=0.4]{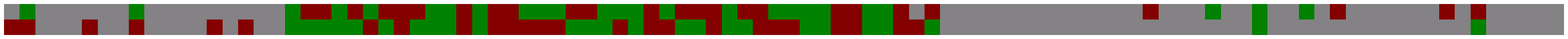}\\
(c) \\
\includegraphics[scale=0.4]{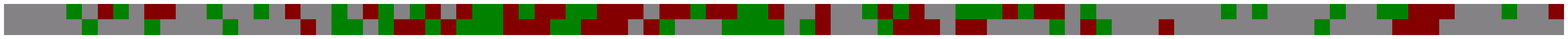}\\
(d) \\
\includegraphics[scale=0.4]{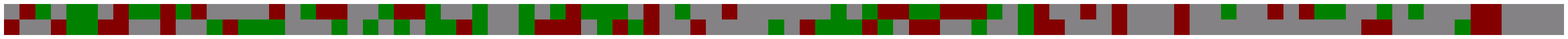}\\
(e) \\
\includegraphics[scale=0.3]{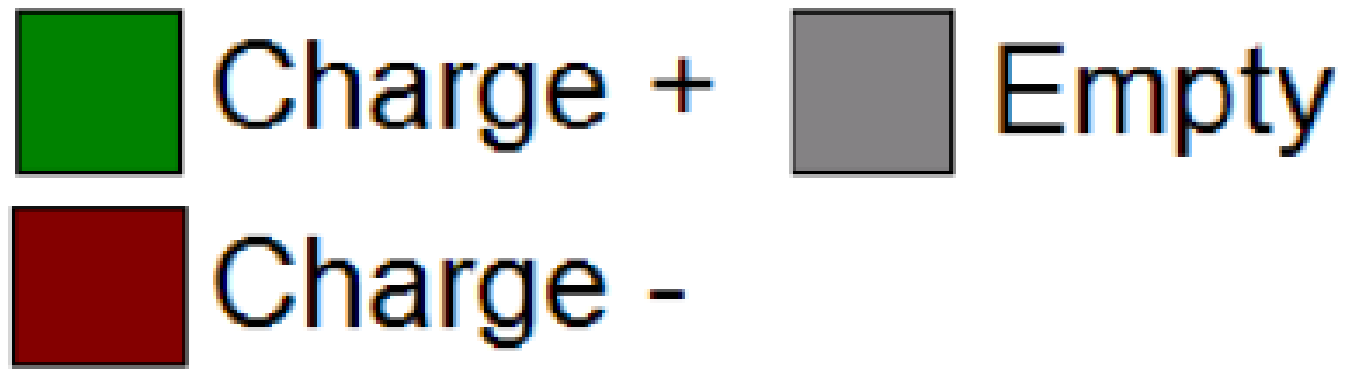}
\end{array}$
\end{center}
\caption{Typical configurations for the gas system in the NESS, for different values of $\gamma$ from the same initial configuration. (a) Initial configuration, (b) $\gamma=0.1$, (c) $\gamma=0.3$, (d) $\gamma=0.5$ and (e) $\gamma=1$ for $t=10000$.}
\label{gasconfig}
\end{figure}

With these dynamical rules, this system evolves with formation of domains for low values of $\gamma$, see figure \ref{gasconfig}-(b) and \ref{gasconfig}-(c). In this regime, the average length of domains grows in time and while the size of the domains remains much smaller than the total size $L$ of the system, the domain size distribution exhibits a dynamic scaling. In the long time limit, the system reaches a non-equilibrium steady state (NESS) where there is only one macroscopic domain. The length of the macroscopic domain depends on $\gamma$, for example for $\gamma=0.1$ it has an approximate size of $L/2$. Additionally, for low values of $\gamma$, this macroscopic domain has not a simple charge distribution and it almost contains no holes. The macroscopic domain is not in equilibrium because there are particles (travelers) which leak out from one end of this domain and travel along the lattice until they reach the other end of this domain, see figure \ref{gasconfig}-(b). In the case of large values of $\gamma$ the system remains homogeneous, i.e., disorganized without domain formation, as we can see in figure \ref{gasconfig}-(e). For intermediate values of $\gamma$, the macroscopic domain is not well formed, it has many holes and it is unstable. In this case there are many travelers and small length domains, see figure \ref{gasconfig}-(d).

In order to obtain quantitative results, we measure the length of a domain by using the coarse grained approximation (CG) defined in \cite{mettetal}. This approximation allows us to map the quasi one-dimensional lattice into a one-dimensional lattice. For any configuration on the $2\times L$ lattice, we construct an effective one-dimensional one, with occupation numbers zero or one on a $L$ sites line, as follows. At each site $i$, we assign $0$ if there are $5$ or less particles in the $10$ sites around it, including the $i$-th column of the original lattice. We assign 1 otherwise, then, we assign $1$ in the $i$-th site of the one-dimensional lattice if there are more particles than holes in the $10$ sites around the $i$-th column in the original lattice \footnote{In our description we take only four neighbor lines around the $i$-th column, two at the left and two at the right. This election is arbitrary however the results are not so sensitive to it. If we take less neighbor lines we improve the measure of the small domains getting worse the measure of big domains and vice versa. Our choice of four neighbor lines allows us a reasonable measure of small and big domains.}. In this simplified description a domain is a simple consecutive sequence of ones and its size is just the length of this string.

\begin{figure}[htp]
\begin{center}
\includegraphics[scale=0.8]{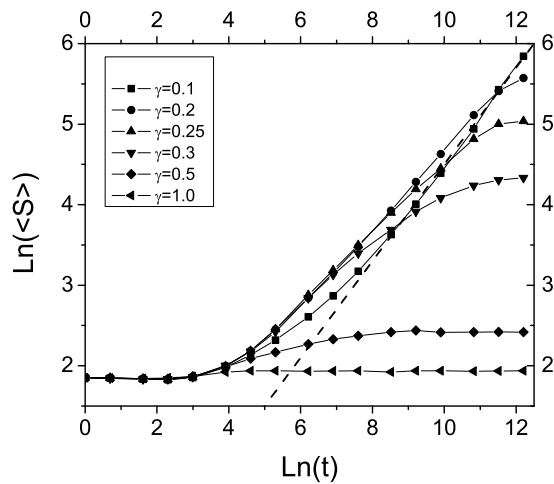}
\end{center}
\caption{Behavior of $\left\langle S(t)\right\rangle$ for different values of $\gamma$, the dashed line is a reference line with slope 0.6. We used $L=1000$.}
\label{gasxprom}
\end{figure}

\begin{figure}[htp]
\begin{center}
\includegraphics[scale=0.8]{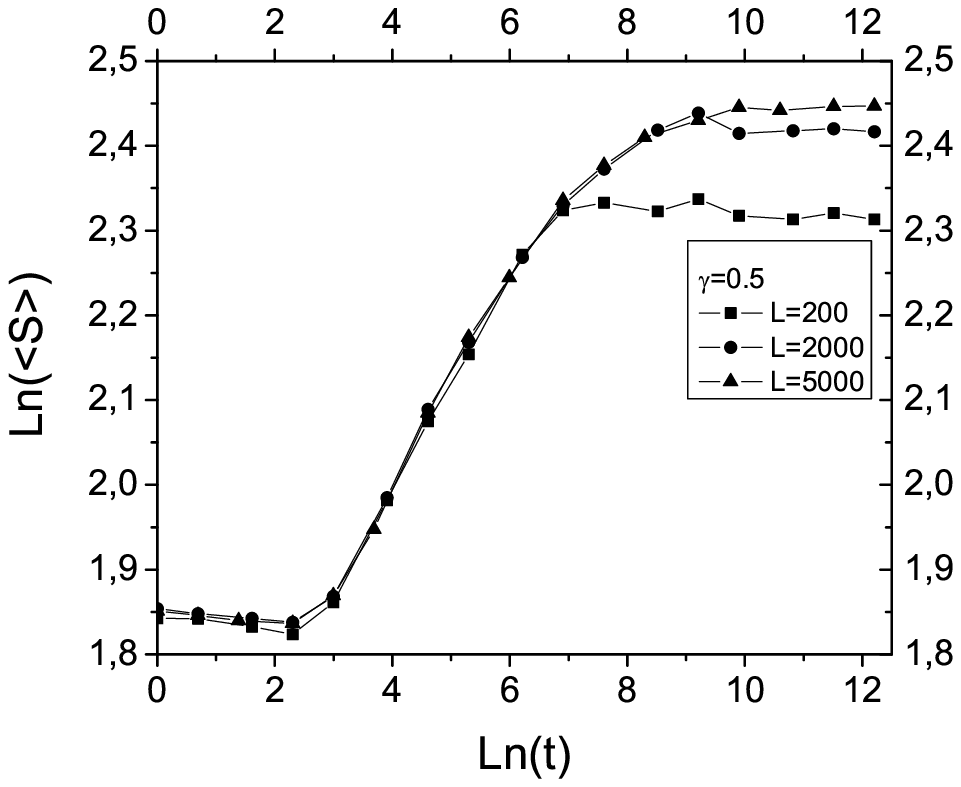}
\end{center}
\caption{Average length of domains $\left\langle S(t)\right\rangle$ in the NESS regime for different values of $L$.}
\label{gasg05ness}
\end{figure}

In figure \ref{gasxprom}, we show the behavior of $\left\langle S(t)\right\rangle$ for six values of $\gamma$. It is evident the NESS behavior in the long time limit when $\left\langle S(t)\right\rangle$ reaches its maximum value which depends on $\gamma$. Additionally, we found that in the NESS regime $\left\langle S(t)\right\rangle$ also depends on the size of the lattice $L$ for all values of $\gamma$, in fact for $\gamma=0.1$ it is well know that $\left\langle S(t)\right\rangle \approx L/2$. In figure \ref{gasg05ness}, we show $\left\langle S(t)\right\rangle$ for $\gamma=0.5$ and different values of $L$. The value of $\left\langle S(t)\right\rangle$ increases with the value of $L$ and we can expect that it reaches its maximum value in the limit $L\rightarrow \infty$ . We can conclude that for finite systems $\left\langle S(t)\right\rangle$, in the NESS, depends on $\gamma$ and $L$.

For $\gamma=0.1$, we found in the scaling regime $\left\langle S(t)\right\rangle\propto t^{0.6}$, this result coincides with the one found in \cite{mettetal}. For lower values of $\gamma$ it seems that there is also a dynamical scaling region whose size decreases when $\gamma$ increases. Naturally, for $\gamma=1$ the systems remains homogeneous. Note that $\left\langle S(t)\right\rangle$ is very different for $\gamma=0.1$ and $\gamma=0.3$ cases,  because for $\gamma=0.3$ there are more travelers than in $\gamma=0.1$, see figure \ref{gasconfig}.

\begin{figure}[htp]
\begin{center}
$\begin{array}{cc}
\includegraphics[scale=0.55]{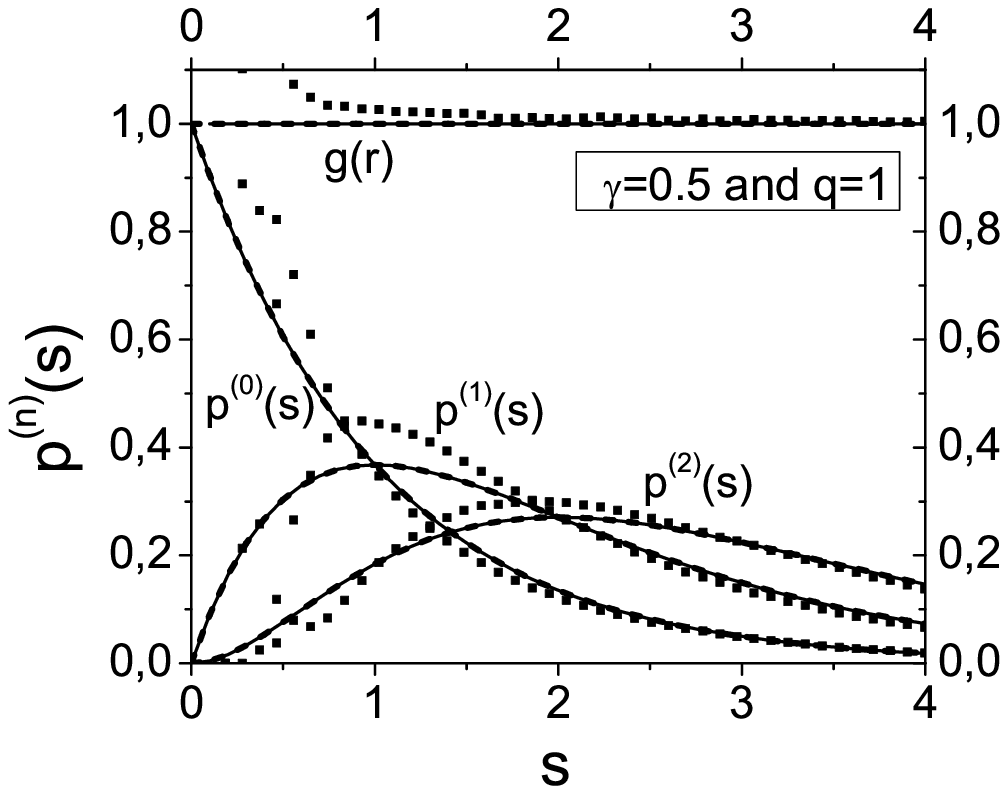}&
\includegraphics[scale=0.55]{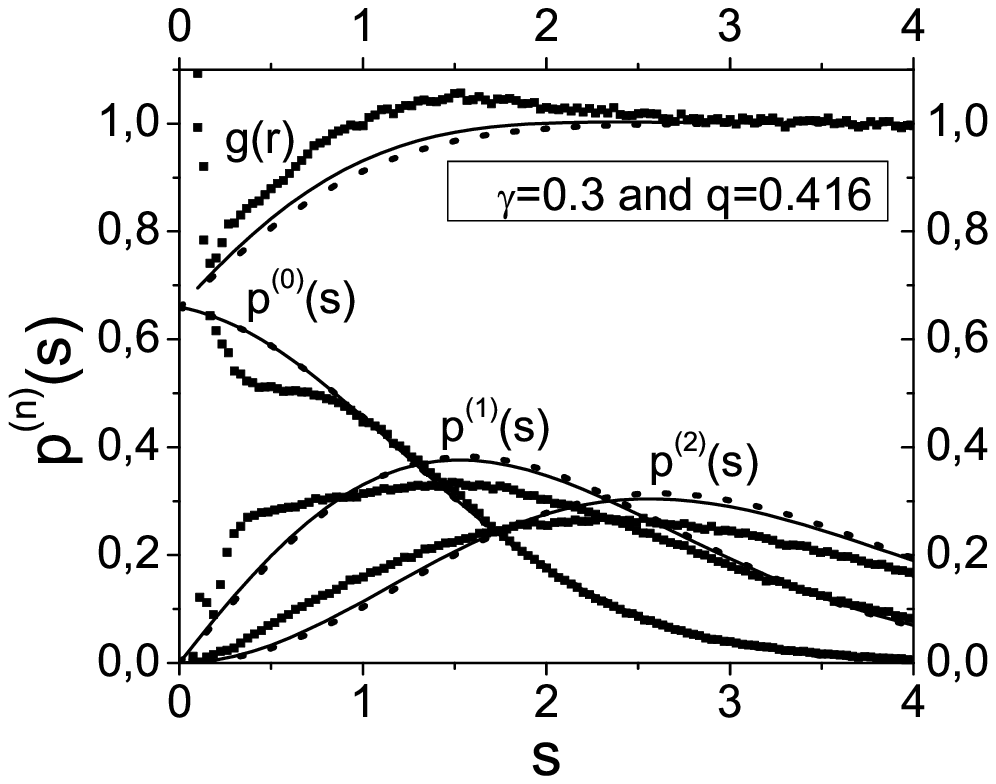}\\
(a) & (b) \\
\includegraphics[scale=0.55]{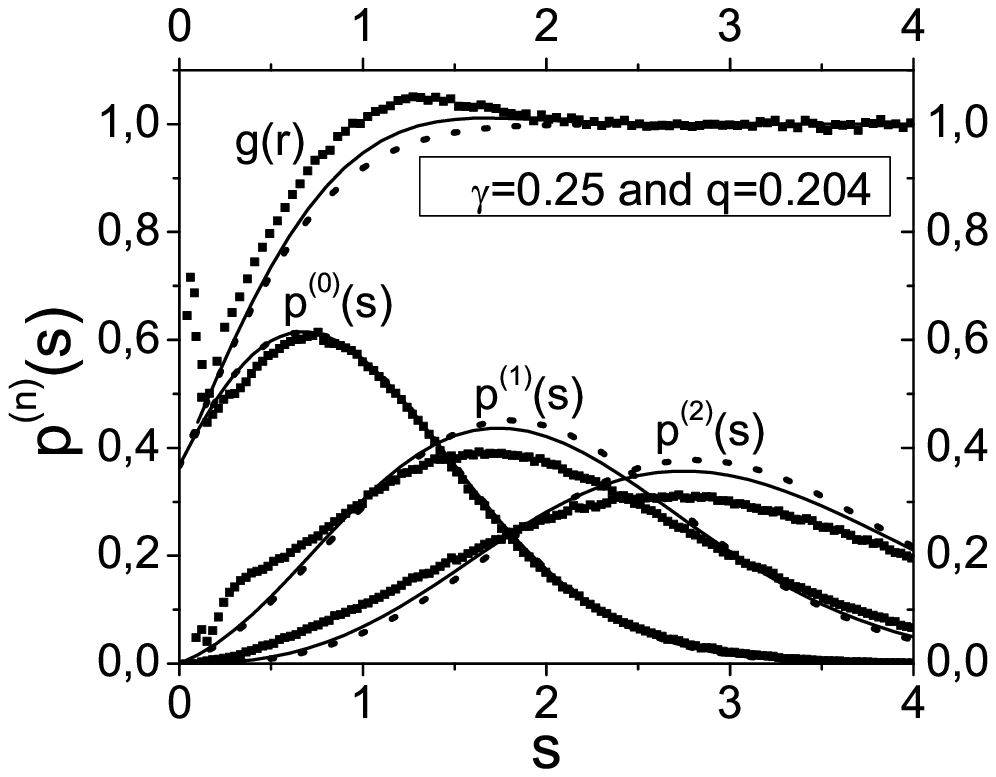}&
\includegraphics[scale=0.55]{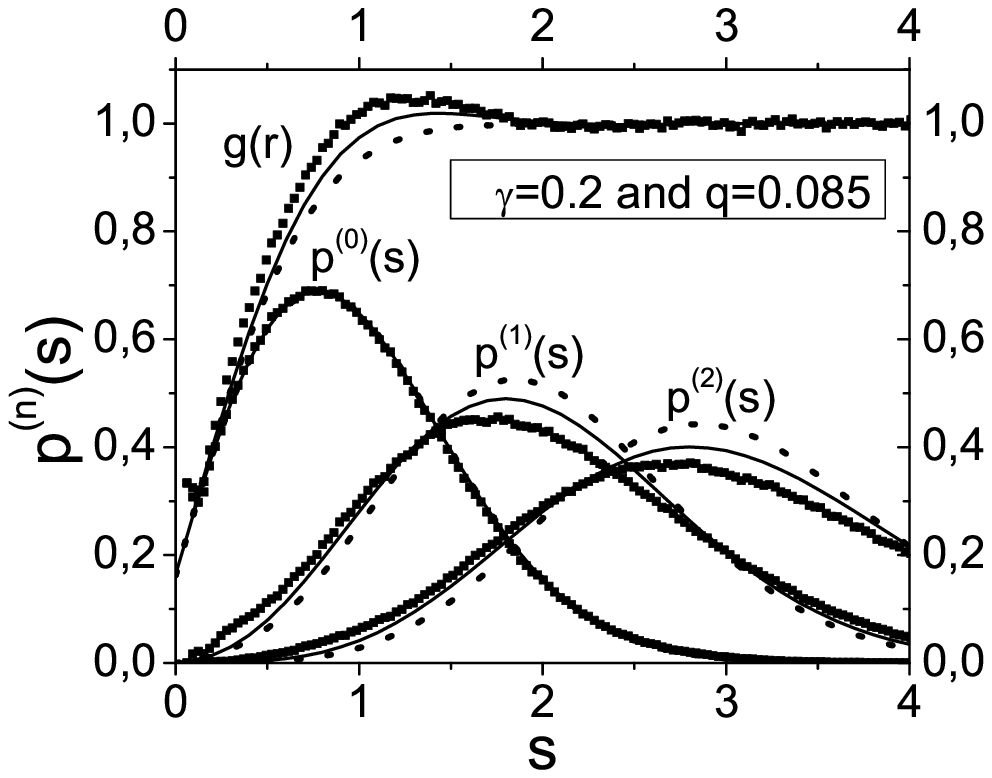}\\
(c) & (d) \\
\end{array}$
$\begin{array}{c}
\includegraphics[scale=0.55]{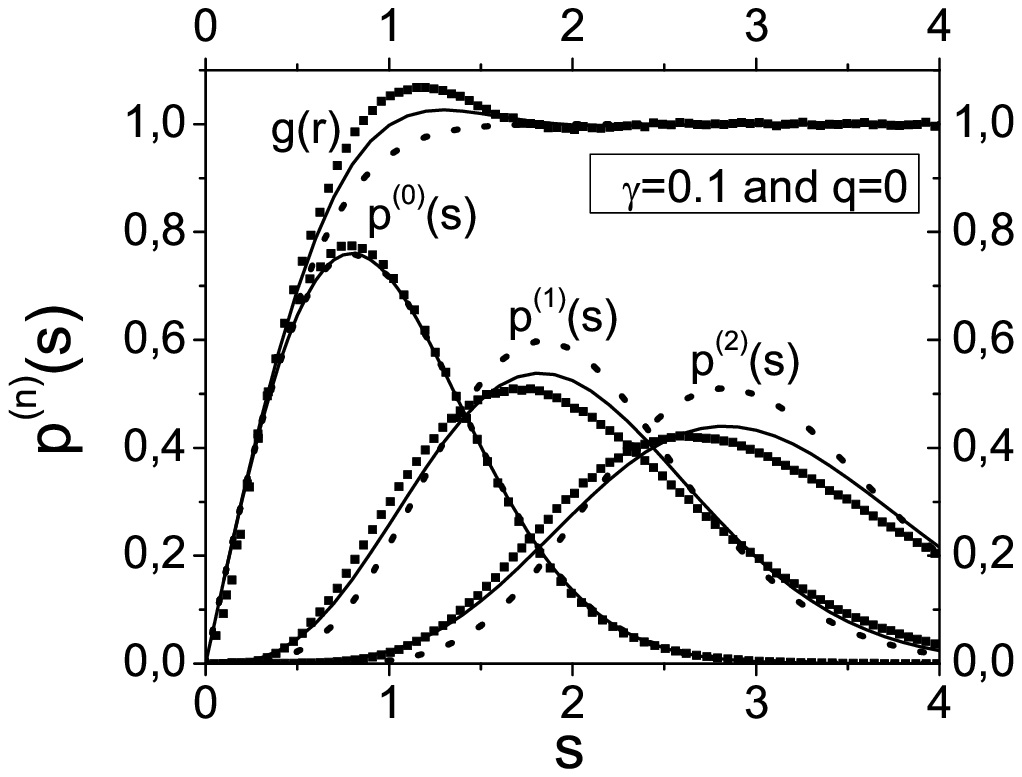}\\
(e)\\
\end{array}$
\end{center}
\caption{Crossover for the gas system. The continuous line corresponds to the BR+IIA model, the dash line corresponds to the GBR model and the square dots correspond to the simulation.}
\label{gastran}
\end{figure}

In figure \ref{gastran}, we show the spacing distribution and the pair correlation functions for the gas system for different values of $\gamma$. As it can be expected, the CG description is not appropriate to measure the length of small domains. However, this method allows us to measure of the length of big domains. For all values of $\gamma$, the nearest neighbor spacing distribution function is well fitted by Berry-Robnik model for high values of $s$. For small values of $\gamma$, $p^{(0)}(s)$ is well described by the Wigner distribution. In the case of $\gamma=1$, the nearest neighbor distribution is described by the Poisson distribution. However, the generalized Berry-Robnik model does not describe the next spacing distributions nor the pair correlation function with enough precision. The differences between the pure coalescing random walk (PCRW for $k=1$) and the gas system was already studied in Ref.~\cite{gonzalez} for the case of $\gamma=0.1$. There the authors found that the independent interval approximation (IIA) is a better model for the gas system than the CRW model. The results that we found for the Berry-Robnik+IIA model are shown in figure \ref{gastran}. We can see that the Berry-Robnik+IIA model is a better approximation for the statistical behavior of the gas for all values of $\gamma$. This fact suggests that, as it happens for $\gamma=1$, the domains in this system are not strongly correlated for all values of $\gamma$ and because of that those correlations can be neglected.

\begin{figure}[htp]
\begin{center}
\includegraphics[scale=0.8]{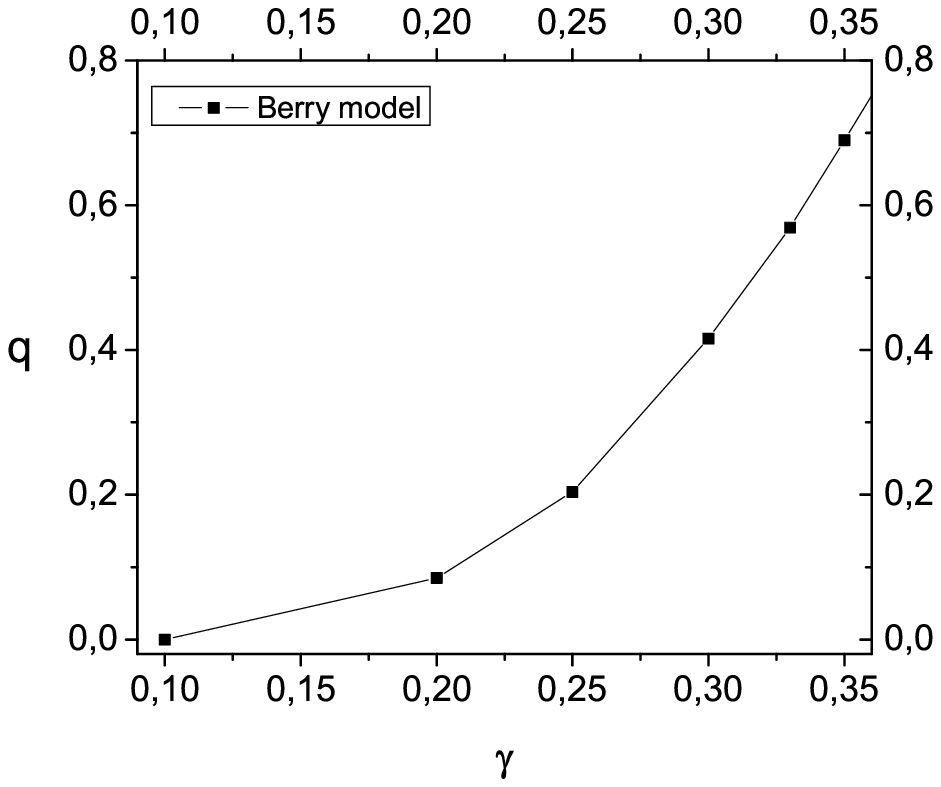}
\end{center}
\caption{Behavior of $q$ in the gas system for different values of $\gamma$.}
\label{gasqvsgama}
\end{figure}

In figure \ref{gasqvsgama}, we show the behavior of $q$ as a function of $\gamma$ in the interval $[0.1,0.35]$. We found that at least for $\gamma>0.4$ the system is in a disorganized state, i.e., its statistical behavior is well described by the Poisson distribution for large values of $s$ and fits give $q\approx1$. In all fits used to find $q$ as function of $\gamma$ we eliminated the regions where the points are highly dispersed, i.e., where the coarse grain method does not measure the length of domains with enough precision.

The numerical results shown in figures \ref{gastran} and \ref{gasqvsgama} were obtained using a $2\times1000$ lattice over $20000$ realizations, the data was taken at three different times $t=2000$, $t=3000$ and $t=4000$. The results shown in figure \ref{gasxprom} were obtained by using the same lattice over $500$ realizations.

\section{Crossover in the spin system}
This system was originally introduced in Ref.~\cite{cornell}. There,  the authors consider a lattice of length $L$ with $L\mu$ spins up (``$+$'') and $L(1-\mu)$ spins down (``$-$'') with $0<\mu<1$. Periodic boundary conditions are imposed. The spin-flip events are:
\begin{enumerate}
	\item $++--$ $\leftrightarrow$ $+-+-$ $\,\,\Delta=4J-E$.
	\item $--++$ $\leftrightarrow$ $-+-+$ $\,\,\Delta=4J+E$.
	\item $++-+$ $\leftrightarrow$ $+-++$ $\,\,\Delta=-E$.
	\item $-+--$ $\leftrightarrow$ $--+-$ $\,\,\Delta=-E$.
\end{enumerate}

The transition probability rate for a process from left to right is $e^{-\frac{\Delta}{T}}$ for $\Delta>0$ and $1$ for $\Delta \leq 0$. The constant $J$ is the nearest neighbor coupling between spins, $E$ is the energy associated to an external field which drives the up (``$+$'') spins to the right and the down (``$-$'') spins to the left, and $T$ the thermal energy (temperature times Boltzmann constant).

In Ref.~\cite{cornell} the authors restrict their study to the regime $T\ll E\ll J$. In this regime the microscopic dynamics of the lattice of spins may be mapped into one for an array domain dynamics, which provides a good approximation in this regime, for more information see \cite{gonzalez,gonzalez3,gonzalez2,cornell,spirin}. With this macroscopic description in Ref.~\cite{gonzalez} the authors show that this system has a statistical behavior very similar to the one of the gas system. Additionally, the system exhibits dynamical scaling behavior and, in the long time limit, the system reaches the NESS, where there are two macroscopic domains which move in opposite directions.
However, this domain model does not allow us study the crossover of the system between organized and disorganized states because it is valid only in the regime $T\ll E\ll J$. To study the crossover regime we must to use the microscopic dynamical rules listed above. In all simulations, we took $E=1$ and $T=1$.

\begin{figure}[htp]
\begin{center}
$\begin{array}{c}
\includegraphics[width=15cm,height=0.3cm]{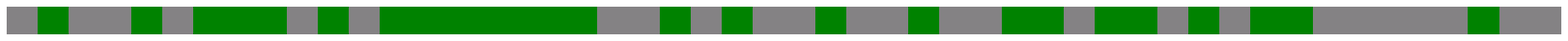}\\
(a)\\
\includegraphics[width=15cm,height=0.3cm]{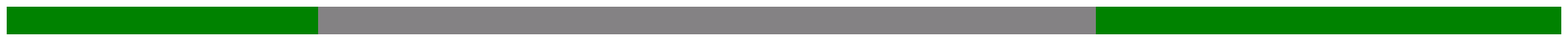}\\
(b)\\
\includegraphics[width=15cm,height=0.3cm]{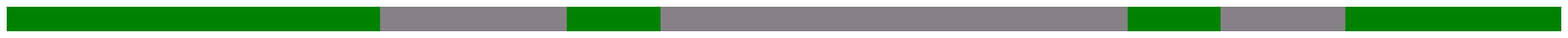}\\
(c) \\
\includegraphics[width=15cm,height=0.3cm]{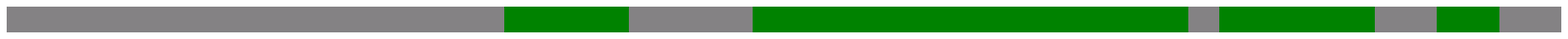}\\
(d) \\
\includegraphics[width=15cm,height=0.3cm]{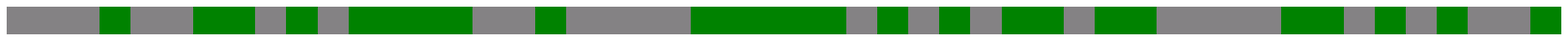}\\
(e) \\
\includegraphics[scale=0.25]{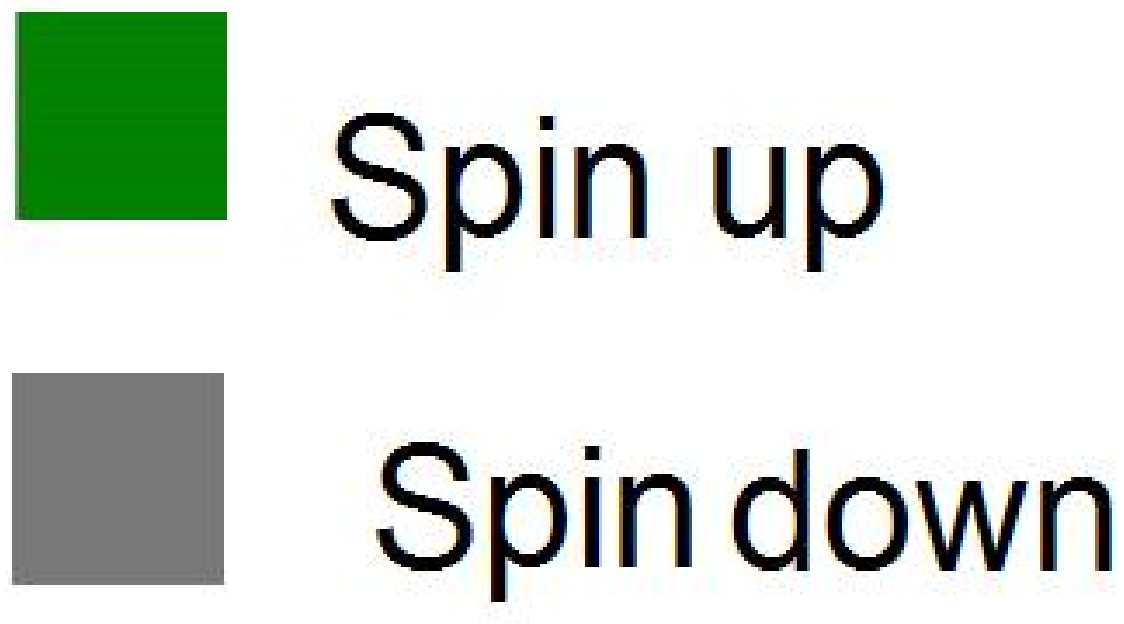}\\
\end{array}$
\end{center}
\caption{Typical configurations for the spin system in the NESS for
  different values of $J$ from the same initial configuration. (a) Initial configuration, (b) $J=2$, (c) $J=1.5$, (d) $J=1$ and (e) $J=0.1$. We use $L=50$, $\mu=0.5$ and $t=100000$.}
\label{spinconfig}
\end{figure}

\begin{figure}[htp]
\begin{center}
\includegraphics[scale=0.8]{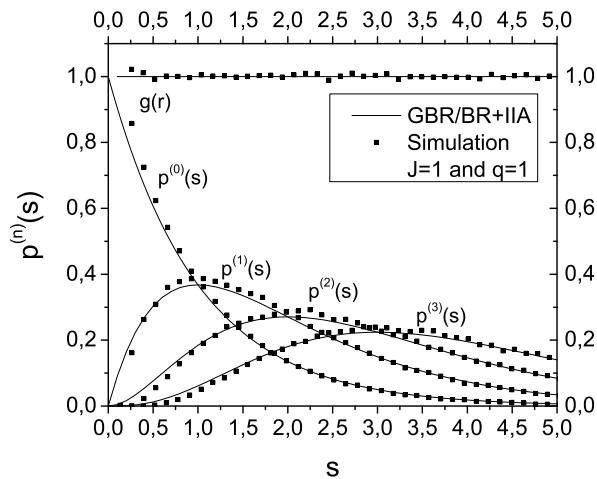}
\end{center}
\caption{Spacing distribution function in the NESS for $J=1$. We took data of three different times $t/L=10,100,1000$ with $L=200$}
\label{spinness}
\end{figure}

In figure \ref{spinconfig}-(b), we can see the asymptotic state for the spin system for large values of $J$ where there are only two stable macroscopic domains and $\left\langle S(t)\right\rangle=L/2$. For $J=0.1$, domain formation is not perceptible, see figure \ref{spinconfig}-(e) which it is almost identically to the disorganized initial state figure \ref{spinconfig}-(a). However, as we will see soon, the average length of domains is different in both cases. In the NESS, for low values of $J$ there is domain destruction and formation, in such way that $\left\langle S(t)\right\rangle$ is constant with a value lower than $L/2$. As we can see in figure \ref{spinness}, for low values of $J$, the statistical behavior of the spin system in the NESS is well described by the Poisson distribution. That means that the system remains in a disorganized state where the average length of domains is bigger than the initial one ($\left\langle S(0)\right\rangle=2$). We can see slight differences near $s=0$, this happens because we used in our simulation a discrete finite lattice. 

\begin{figure}[htp]
\begin{center}
\includegraphics[scale=0.8]{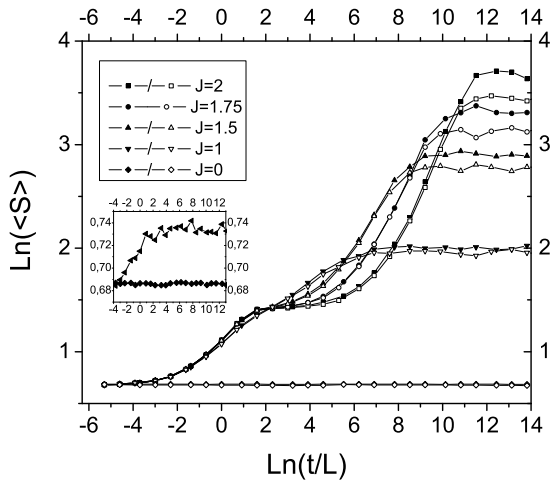}
\end{center}
\caption{Average length for different values of $J$. Open symbols: $L=100$ and filled symbols: $L=200$. In the inset we compare $J=0$ and $J=0.1$ for $L=200$ in order to verify the domain formation for small values of $J$.}
\label{spinxprom}
\end{figure}

\begin{figure}[htp]
\begin{center}
\includegraphics[scale=0.8]{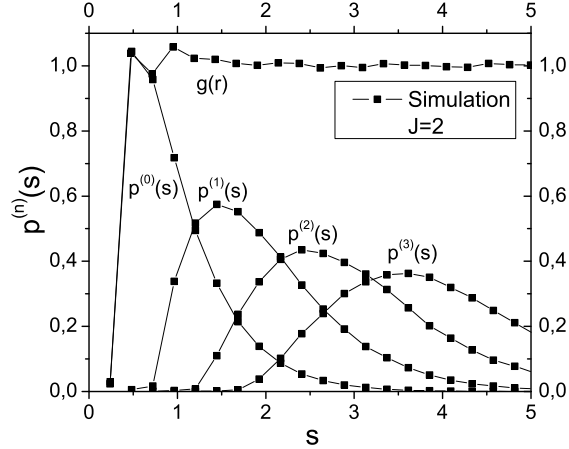}
\end{center}
\caption{Statistical behavior of the metastable state in the spin system for $J=2$. We took data at tree different times: $T/L=10,20,30$.}
\label{meta}
\end{figure}

In figure \ref{spinxprom}, we can see the average length of domains
$\left\langle S(t)\right\rangle$ as a function of time $t$. In this
case the behavior of $\left\langle S(t)\right\rangle$ is different
from the one found in the PCRW and in the gas system. For large values
of $J$, the spin system reaches metastable states where $\left\langle
S(t)\right\rangle$ is almost constant. Note that in all cases, the
metastable state starts when the domain density is $1/4$, i.e., when
the system, in the statistical average sense, is filled with domains
of the form $++--++--$. This is a consequence of the microscopical
dynamical rules that we use, because they take into account interaction
among four neighbors spins. For large enough values of $J$, these
domains have a low probability of destruction setting the system into
a metastable state. In figure \ref{meta}, we show the spacing
distribution function in the metastable states, the level repulsion is
present. After some time, $\left\langle S(t)\right\rangle$ starts to
grow again in time. In the PCRW and in the gas systems the metastable
states are not present. Once the metastable states ends, the domain
formation in the system continues in such way that the average size of
domains increases again. The length in time of these metastable states
depends on the value of $J$ (with $E=T=1$). In fact for $J=1$ this
region is absent but for bigger values of $J$ its size
increases. Nevertheless, $\left\langle S(t)\right\rangle$ does not
seem to depend on the length of the system in the scaling regime, it
is the same for $L=100,200,500$, the differences arises near to the
NESS naturally \footnote{The case $L=500$ was verified but is not
  included in the figure.}. In figure \ref{spinxprom}, for $J=2$, the
system seems to reach another metastable state where $\left\langle
S(t)\right\rangle\approx 40$ for $N=200$ and $\left\langle
S(t)\right\rangle\approx 30$ for $N=100$. For $J=2$, the scaling
region has a considerable size but this region for lower values of $J$
is smaller. From figure \ref{spinxprom} it is evident that the value
of $\left\langle S(t)\right\rangle$ in the NESS regime depends on the
size of the system, as it also happens in the gas system. In the inset
of figure \ref{spinxprom}, we compare $\left\langle S(t)\right\rangle$
for $J=0$ and $J=0.1$, in the last case there is domain formation and
it seems that only for $J=0$ there is not domain formation.

\begin{figure}[htp]
\begin{center}
$\begin{array}{cc}
\includegraphics[scale=0.6]{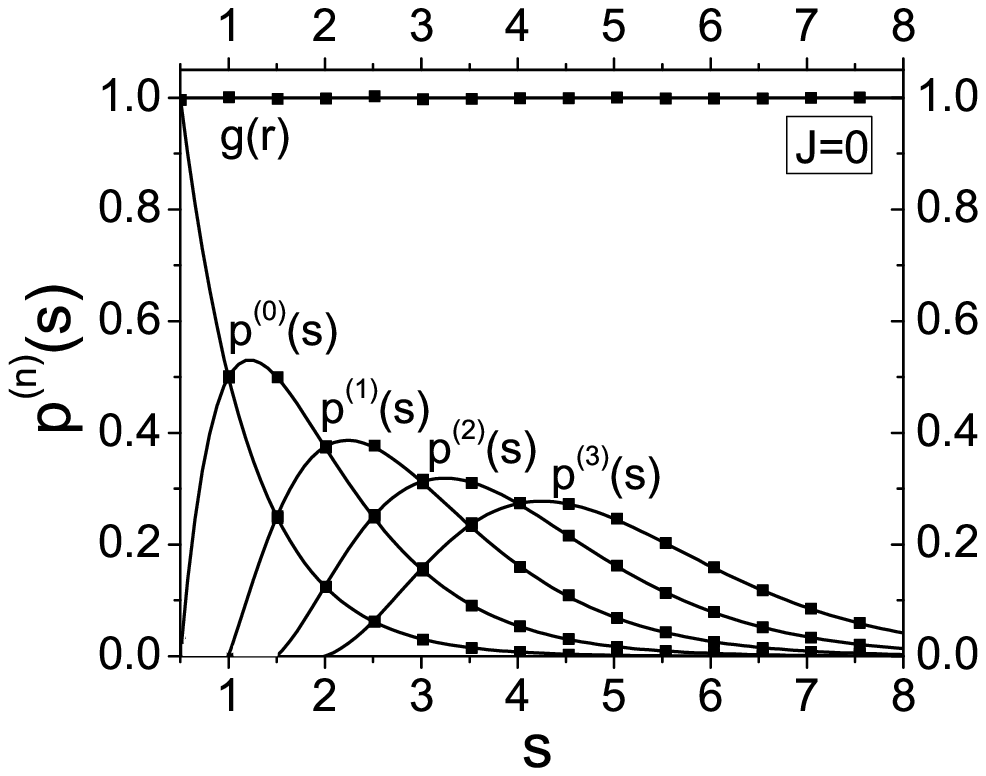}&
\includegraphics[scale=0.6]{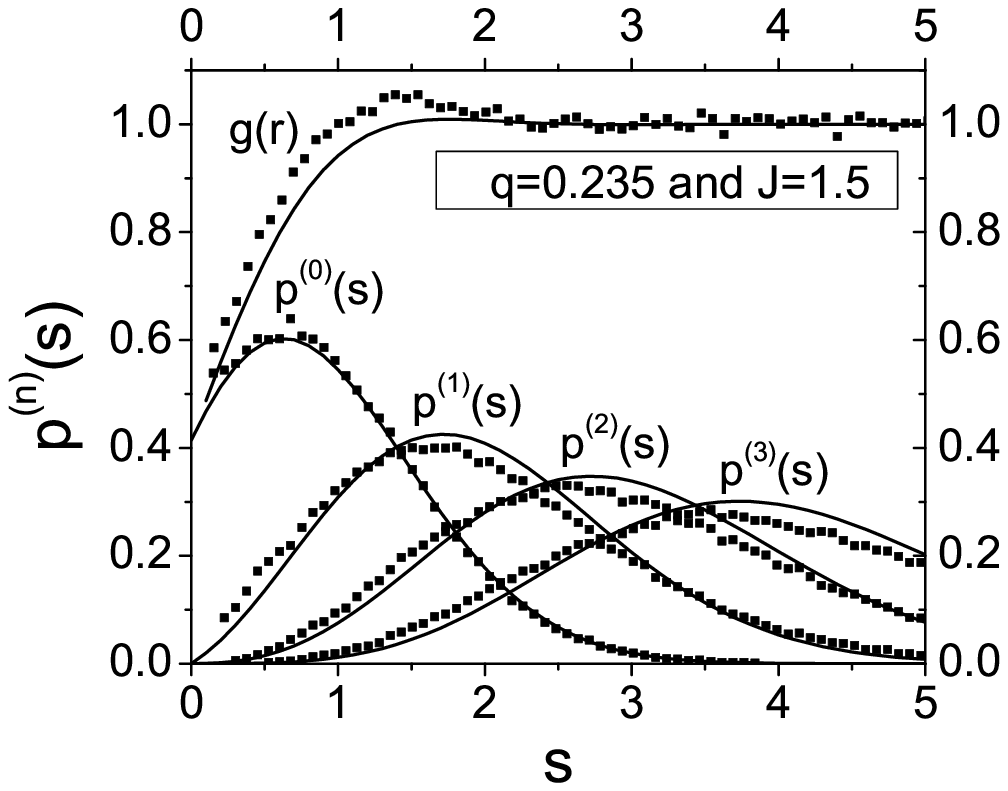}\\
(a) & (b) \\
\includegraphics[scale=0.6]{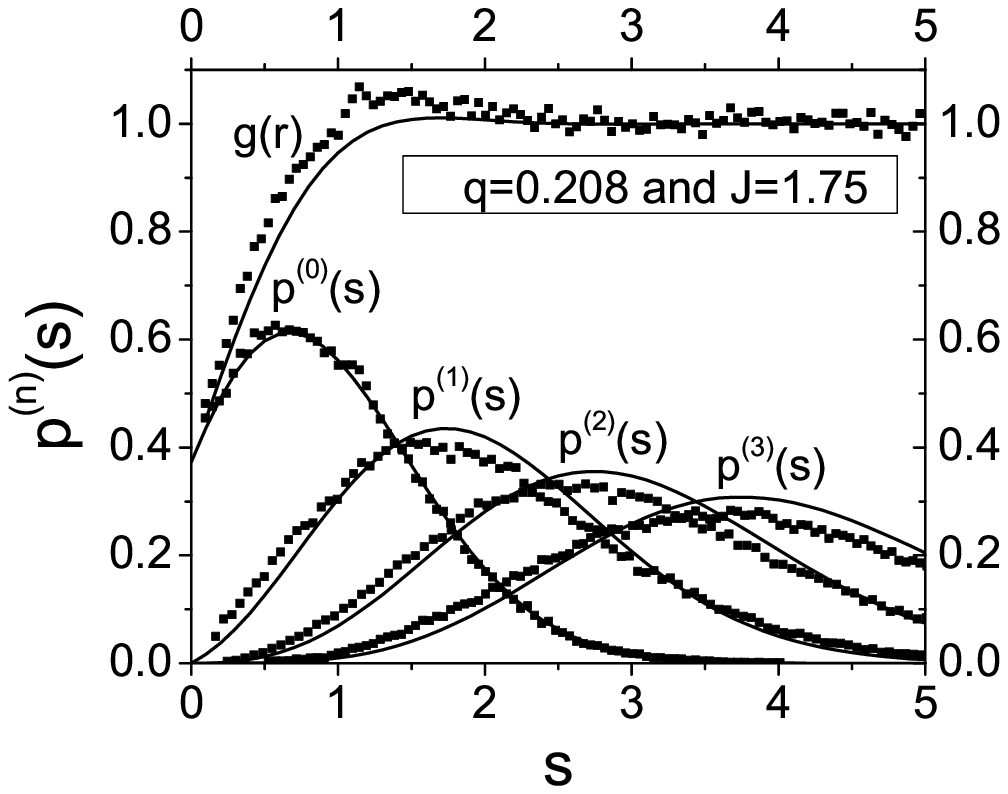}&
\includegraphics[scale=0.6]{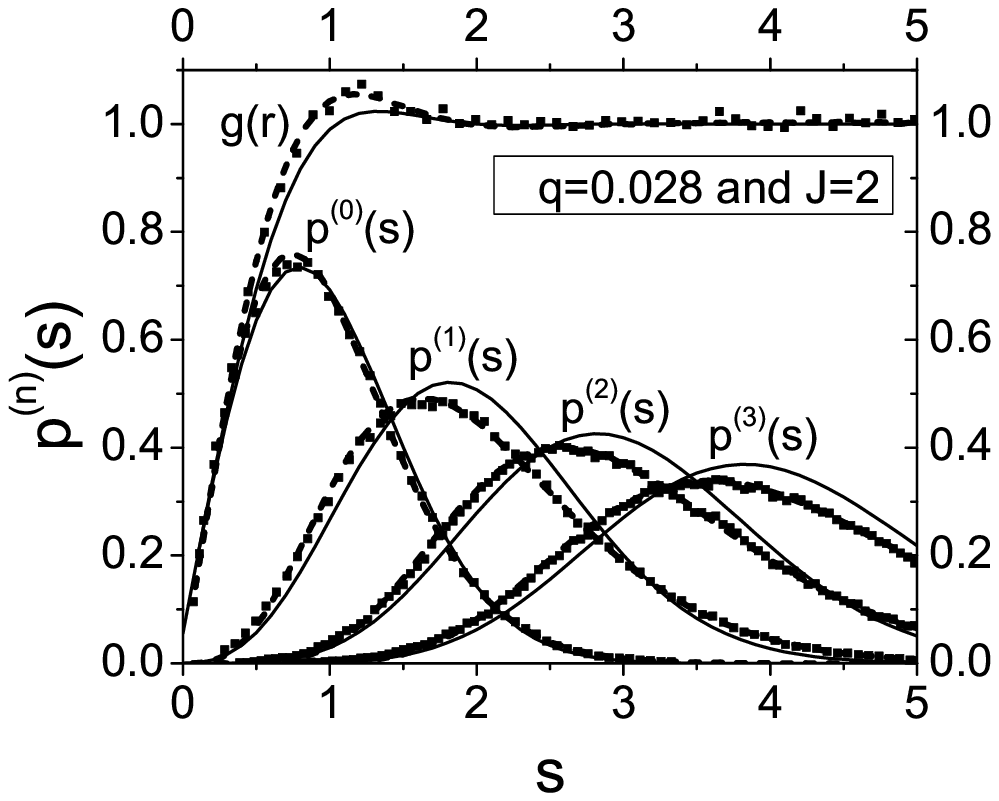}\\
(c) & (d) \\
\end{array}$
\end{center}
\caption{Crossover for the spin system.}
\label{spintran}
\end{figure}

As it happens in the gas system, the statistical behavior of this model is better described by the Berry-Robnik+IIA than the generalized Berry-Robnik model. In figure \ref{spintran}, we show the statistical behavior of the system for different values of $J$. For $J=0$, the system evolves in time remaining disorganized, as we expected. However, we cannot compare directly its spacing distributions with our analytical model because of the effects of the discrete finite lattice that we use in our simulations. In fact, as we can see in figure \ref{spintran}, $p^{(0)}(1/2)=1$ instead of $p^{(0)}(0)=1$ as it is predicted by the continuous Poisson distribution. This result can be understood if we remember that, in a lattice, the lowest value for which the nearest spacing distribution is defined, is $s=\rho(t)$, with $\rho(t)$ the density of particles. In our case, $\rho(t)=1/2$ for $J=0$ in each spin species. This technical problem is usually solved taking low density values as we did it in the PCRW, where we took $\rho(0)=1/10$. However, we cannot do that in the spin system. We have to take small lattices because the simulation is very intensive  and low densities give us a poor statistic. This is only a technical problem and we can be sure that in the limit of low densities and small values of $J$, the system is well described by the continuous Poisson distribution. To corroborate this, we compare the results of our simulation for $J=0$ with the discrete version of the Poisson distribution ($p^{(0)}(s)=(1/2)^{2s-1}$,  $p^{(1)}(s)=(2s-1)(1/2)^{2s-1}$, etc.). 

For $J=2$ the system sets into an organized state where there is domain formation. In this case, $p^{(0)}(s)$ is well fitted by the Wigner distribution. For $J=2$, the results of our microscopic simulation coincides with the results of the macroscopic array domains simulation in the regime $T\ll E \ll J$. For intermediate values, the system shows a mixed state. If we do not take into account the discrete finite lattice, for all values of $J$, $p^{(n)}(s)$ is well described by the Berry-Robnik+IIA model. This suggest that for this system the correlation between domains are not strong as it happen in the gas system. For low densities and values of $J$ near 1, we expect that the generalized Berry-Robnik and the Berry-Robnik+IIA models are a good approximation for higher spacing distribution functions ($n>1$) and for the pair correlation function because in this regime the correlation between domains can be neglected.

In our numerical simulations we took $\mu=0.5$ and $2000$ realizations. The measures were taken at different times using two lattices: $L=500$ and $L=200$, in order to confirm the scaling property. For $J=2$, we took $t/L=10000$ and $t/L=4000$ for $L=500$ and $L=200$. For $J=1.75$, we took $t/L=9000$ for $L=500$ and $t/L=6000,7500,9000$ for $L=200$. For $J=1.5$, we took $t/L=2000$ for $L=500$ and $t/L=1500,2000$ for $L=200$. For $J=0$, the system remains disorganized and the time when the data is taken is irrelevant.

\section{Conclusion}
In all systems considered in this paper, the original Berry-Robnik model gives us a good fit for the nearest neighbor distribution $p^{(0)}(s)$, i.e., $p^{(0)}(s)$ can be approximated by the uncorrelated superposition of one Wigner and one Poisson distribution functions. The biggest differences in the fits are given when the densities of both sequences have similar values. We calculate the next spacing distribution functions by using two methods, the GBR and the BR+IIA. We found that the GBR model gives us a good approximation for the statistical behavior of the PCRW and the BR+IIA model is a good approximation for the gas and spin systems. This fact suggest that the correlation between domains in the gas and spin system are not strong and can be neglected in a first approximation but in the PCRW the correlations between particles cannot be neglected. 

Our analytical models are simple and allow us to have a quantitative
measure of the degree of order/disorder of the systems through the
parameter $q$. For $q=1$ the system is completely disordered and
$p^{(0)}(s)$ is described by the Poisson distribution, that means that
the system is homogeneous, in the statistical average sense. For $q=0$
the system is organized and $p^{(0)}(s)$ is described by the Wigner
distribution. In the particular case of the PCRW $q=0$ implies that
there is an effective repulsion between particles and in the case of
the gas and spin systems means domain formation. Unfortunately, the
information about the interaction between particles/domains cannot be
extracted easily from the statistical behavior in non-equilibrium
systems. However, by using the BR+IIA model it is possible to
calculate in an approximate way the interaction potential between
domain borders for the gas and spin systems, as
follows. Comparing~(\ref{pnxn}) with a Boltzmann factor with an
inverse temperature $\beta=1$, is straightforward to show that, under
the BR+IIA approximation, the statistics of the domain borders is
approximately equivalent to a system with $N$ particles which interact
according to the potential
\begin{equation}\label{viia}
V_N(x_1,\cdots,x_N) = \sum^{N}_{i=1} q \left(x_{i+1}-x_i\right)-\sum^{N}_{i=1}\mathrm{ln}\left[f(x_i,x_{i+1})\right],
\end{equation}
where 
\begin{eqnarray}
f(x_i,x_{i+1})=q^2\mathrm{erfc}\left(\frac{\sqrt{\pi}}{2}(1-q)s\right)+\left(2 q(1-q)+\frac{\pi}{2}(1-q)^3 s\right)e^{-\frac{\pi}{4}(1-q)^2 s^2},
\end{eqnarray}
and $s\equiv x_{i+1}-x_i$. Thus, the statistics of the domain edges of the gas and spin systems in their crossover regimes is approximately equivalent to a statistical equilibrium system of particles on a circle interacting through the nearest neighbor pair potential (\ref{viia}). Note that for $q=0$, Eq.~(\ref{viia}) reduces to
\begin{equation}\label{viia1}
V_N(x_1,\cdots,x_N) =
\sum^{N}_{i=1}\left[
\frac{\pi}{4}\left(x_{i+1}-x_i\right)^2
-\ln\left(\frac{\pi}{2}\left(x_{i+1}-x_i\right)\right)
\right]\,,
\end{equation}
as we can expect from Ref.~\cite{gonzalez}. Naturally, for $q=1$ the interaction potential is a constant and the interaction force between domain edges  vanishes. 

In the crossover between order and disorder all systems lose their level repulsion properties, i.e., the correlations between domains/particles decrease. 
In the PCRW, the correlations between particles arises from the coalescence reaction for $k>0$, in the gas case, correlations arise from the mutual obstruction of particles for $\gamma<1$. For the spin system, taking $E=1$ and $T=1$, the crossover depends only on $J$ and the correlations arise for $J>0$.

The gas and spin systems have a similar statistical behavior in the crossover between organized and disorganized states but $\left\langle S(t)\right\rangle$ is very different in both cases. The metastable regions that we found in the spin system are not present in the gas system nor in the PCRW. In fact metastable regions are not predicted by the macroscopic dynamical rules used in Ref.~\cite{cornell}. In the PCRW there is a time $\tau_1$ where the interaction goes unnoticed which depends on the parameter $k$. In the gas an spin systems it seems that this time does not depend on the parameters $\gamma$ and $J$ respectively.

Finally, for the PCRW, we found that the parameter $q\equiv
q(\tau,k)$, which characterize the crossover, can be scaled in a
function of a single argument $\tilde{\tau}$ by making the change of
variable $\tilde{\tau}=\tau k^2/(1-k)^2$. The scaling factor is
proportional to the crossover time between intermediate regime and
long time regime, $\tau_2$.

\section*{Acknowledgments}
The authors thank F.~van Wijland for valuable
comments and observations. This work was partially supported by an ECOS Nord/COLCIENCIAS action
of French and Colombian cooperation and by Comit\'e de Investigaciones
y Posgrados, Facultad de Ciencias, Universidad de los Andes.


\begin{thebibliography}{99}
\bibitem{berry} M.~V.~Berry and M.~Robnik, Semiclassical level spacings when regular and 
chaotic orbits coexist, J.~Phys.~A: Math. Gen.~\textbf{17}, 2413--2421 (1984).
\bibitem{amiet} P.~Jacquod and J.~P.~Amiet, Evidence for the validity of the Berry-Robnik surmise in a periodically pulsed spin system, J.~Phys.~A: Math. Gen.~\textbf{28}, 4799--4811 (1995).
\bibitem{lopac} V.~Lopac, S.~Brant and V.~Paar, Level density fluctuations and characterization of chaos in the realistic model spectra for odd-odd nuclei, Z.~Phys.~A.~\textbf{356}, 113--118 (1996).
\bibitem{ben}D.~ben-Avraham and S.~Havlin. Diffusion and reactions in fractals and disordered systems, Cambridge University Press (2000).
\bibitem{gonzalez}D.~L.~González and G.~Téllez, Statistical behavior of domain systems, 
Phys. Rev. E \textbf{76}, 011126 (2007). 
\bibitem{gonzalez3}D. L. González and G. Téllez, Is The Nearest Neighbor Distribution Enough to Describe The Statistical Behavior Of A Domain System?, To be published in the Traffic granular flow 2007. 
\bibitem{ben2} D.~ben-Avraham and É.~Brunet, On the relation between one-species diffusion-limited coalescence and annihilation in one dimension, J. Phys. A: Math. Gen. ~\textbf{38}, 3247--3252 (2005).
\bibitem{gonzalez2}D. L. González and G. Téllez, Wigner domains for domain systems, J. Stat. Phys. \textbf{132}, 187--205 (2008). 
\bibitem{ben0} D.~Zhong and D.~Ben-Avraham. Diffusion-limited coalescence with finite reaction rates in one dimension. J. Phys. A: Math. Gen. \textbf{28} 33--44 (1995). 
\bibitem{priv}V. Privman and C. R. Doering, Crossover from rate-equation to diffusion-controlled kinetics in two particle coagulation, Phys. Rev. E \textbf{48}, 846--851 (1993). 
\bibitem{mettetal} J.~Mettetal, B.~Schmittmann and R.~Zia, Coarsening
dynamics of a quasi one-dimensional driven lattice gas, Europhysics
Lett.~\textbf{58}, 653--659 (2002).
\bibitem{cornell} S.~J.~Cornell and A.~J.~Bray, Domain growth in a
  one-dimensional driven diffusive system, Phys.~Rev.~E \textbf{54},
  1153--1160 (1996).
\bibitem{spirin} V.~Spirin, P.~L.~Krapivsky and S.~Redner, Coarsening
  in a driven Ising chain with conserved dynamics, Phys.~Rev.~E
  \textbf{60}, 2670--2676 (1999).
\bibitem{alemany} P.~A.~Alemany and D.~ben-Avraham, Inter-particle
  distribution functions for one-species diffusion-limited
  annihilation, $A+A\to0$, Phys.~Lett.~A.~\textbf{206}, 18--25 (1995).
\bibitem{derrida} B.~Derrida, V.~Hakim and R.~Zeitak, Persistent spins
  in the linear diffusion approximation of phase ordering and zeros of
  stationary gaussian processes, Phys.~Rev.~Lett.~\textbf{77},
  2871--2874 (1996).
\bibitem{majumdar} S.~N.~Majumdar, C.~Sire, A.~J.~Bray and
  S.~J.~Cornell, Nontrivial exponent for simple diffusion,
  Phys.~Rev.~Lett.~\textbf{77}, 2867--2870 (1996).
\bibitem{Krap-Naim-PRE} P.~L.~Krapivsky and E.~Ben-Naim, Domain
  statistics in coarsening systems, Phys.~Rev.~E \textbf{56},
  3788--3798 (1997).
\bibitem{Krap-Naim-JSP} E.~Ben-Naim and P.~L.~Krapivsky, Domain number
  distribution in the nonequilibrium Ising model,
  J.~Stat.~Phys.~\textbf{93}, 583--601 (1998).
\bibitem{salsburg} Z.~W.~Salsburg, R.~W.~Zwanzig and
  J.~G.~~Kirkwood, 
  Molecular distribution functions in a one-dimensional fluid,
  J.~Chem.~Phys.~\textbf{21}, 1098--1107 (1953).

\end{thebibliography}
\end{document}